\documentclass[12pt]{article}
\usepackage{latexsym}
\usepackage{amssymb}
\usepackage{a4}
\usepackage{graphics,graphicx}
\usepackage{tabularx}
\usepackage{amsfonts}
\usepackage{amsmath}
\usepackage{subfig}
\usepackage[T1]{fontenc} % output font encoding
\usepackage[latin1]{inputenc} % output font encoding
\usepackage{epstopdf}

\usepackage{color}

\newcommand{\footmsg}[1]{%
  \let\temp\thempfn%
  \def\thempfs{}
  \footnotetext{#1}
  \let\tempfn\temp}

\begin{document}

%Definitions: general
\newcommand{\singlespace} {\baselineskip=12pt
\lineskiplimit=0pt \lineskip=0pt }
\def\ds{\displaystyle}

%Definitions: equations
\newcommand{\beq}{\begin{equation}}
\newcommand{\eeq}{\end{equation}}
\newcommand{\lb}{\label}
\newcommand{\beqar}{\begin{eqnarray}}
\newcommand{\eeqar}{\end{eqnarray}}
\newcommand{\barr}{\begin{array}}
\newcommand{\earr}{\end{array}}

\newcommand{\jump}{\parallel}

\def\c{{\circ}}

\newcommand{\Ehat}{\hat{E}}
\newcommand{\That}{\hat{\bf T}}
\newcommand{\Ahat}{\hat{A}}
\newcommand{\chat}{\hat{c}}
\newcommand{\shat}{\hat{s}}
\newcommand{\khat}{\hat{k}}
\newcommand{\muhat}{\hat{\mu}}
\newcommand{\mc}{M^{\scriptscriptstyle C}}
\newcommand{\mei}{M^{\scriptscriptstyle M,EI}}
\newcommand{\mec}{M^{\scriptscriptstyle M,EC}}

\newcommand{\hbeta}{{\hat{\beta}}}
\newcommand{\rec}[2]{\left( #1 #2 \ds{\frac{1}{#1}}\right)}
\newcommand{\rep}[2]{\left( {#1}^2 #2 \ds{\frac{1}{{#1}^2}}\right)}
\newcommand{\derp}[2]{\ds{\frac {\partial #1}{\partial #2}}}
\newcommand{\derpn}[3]{\ds{\frac {\partial^{#3}#1}{\partial #2^{#3}}}}
\newcommand{\dert}[2]{\ds{\frac {d #1}{d #2}}}
\newcommand{\dertn}[3]{\ds{\frac {d^{#3} #1}{d #2^{#3}}}}

\def\bob{{\, \underline{\overline{\otimes}} \,}}

\def\ob{{\, \underline{\otimes} \,}}
\def\scalp{\mbox{\boldmath$\, \cdot \, $}}
\def\gdp{\makebox{\raisebox{-.215ex}{$\Box$}\hspace{-.778em}$\times$}}

\def\daa{\makebox{\raisebox{-.050ex}{$-$}\hspace{-.550em}$: ~$}}

\def\mK{\mbox{${\mathcal{K}}$}}
\def\cK{\mbox{${\mathbb {K}}$}}

%Definitions: integrals
\def\Xint#1{\mathchoice
   {\XXint\displaystyle\textstyle{#1}}%
   {\XXint\textstyle\scriptstyle{#1}}%
   {\XXint\scriptstyle\scriptscriptstyle{#1}}%
   {\XXint\scriptscriptstyle\scriptscriptstyle{#1}}%
   \!\int}
\def\XXint#1#2#3{{\setbox0=\hbox{$#1{#2#3}{\int}$}
     \vcenter{\hbox{$#2#3$}}\kern-.5\wd0}}
\def\ddashint{\Xint=}
\def\fpint{\Xint=}
\def\dashint{\Xint-}
\def\cpvint{\Xint-}
\def\intl{\int\limits}
\def\cpvintl{\cpvint\limits}
\def\fpintl{\fpint\limits}
\def\ointl{\oint\limits}

\def\bA{{\bf A}}
\def\ba{{\bf a}}
\def\bB{{\bf B}}
\def\bb{{\bf b}}
\def\bc{{\bf c}}
\def\bC{{\bf C}}
\def\bD{{\bf D}}
\def\bd{{\bf d}}
\def\bE{{\bf E}}
\def\be{{\bf e}}
\def\bbf{{\bf f}}
\def\bF{{\bf F}}
\def\bG{{\bf G}}
\def\bg{{\bf g}}
\def\bi{{\bf i}}
\def\bH{{\bf H}}
\def\bK{{\bf K}}
\def\bL{{\bf L}}
\def\bM{{\bf M}}
\def\bN{{\bf N}}
\def\bn{{\bf n}}
\def\bm{{\bf m}}
\def\b0{{\bf 0}}
\def\bo{{\bf o}}
\def\bX{{\bf X}}
\def\bx{{\bf x}}
\def\bP{{\bf P}}
\def\bp{{\bf p}}
\def\bQ{{\bf Q}}
\def\bq{{\bf q}}
\def\bR{{\bf R}}
\def\bS{{\bf S}}
\def\bs{{\bf s}}
\def\bT{{\bf T}}
\def\bt{{\bf t}}
\def\bU{{\bf U}}
\def\bu{{\bf u}}
\def\bv{{\bf v}}
\def\bw{{\bf w}}
\def\bW{{\bf W}}
\def\by{{\bf y}}
\def\bz{{\bf z}}
\def\bk{{\bf k}}

\def\Id{{\bf I}}
\def\bxi{\mbox{\boldmath${\xi}$}}
\def\balpha{\mbox{\boldmath${\alpha}$}}
\def\bbeta{\mbox{\boldmath${\beta}$}}
\def\bchi{\mbox{\boldmath${\chi}$}}
\def\bepsilon{\mbox{\boldmath${\epsilon}$}}
\def\bvarepsilon{\mbox{\boldmath${\varepsilon}$}}
\def\bomega{\mbox{\boldmath${\omega}$}}
\def\bphi{\mbox{\boldmath${\phi}$}}
\def\bsigma{\mbox{\boldmath${\sigma}$}}
\def\bfeta{\mbox{\boldmath${\eta}$}}
\def\bDelta{\mbox{\boldmath${\Delta}$}}
\def\btau{\mbox{\boldmath $\tau$}}

\def\tr{{\rm tr}}
\def\dev{{\rm dev}}
\def\div{{\rm div}}
\def\Div{{\rm Div}}
\def\Grad{{\rm Grad}}
\def\grad{{\rm grad}}
\def\Lin{{\rm Lin}}
\def\Sym{{\rm Sym}}
\def\Skw{{\rm Skew}}
\def\abs{{\rm abs}}
\def\Re{{\rm Re}}
\def\Im{{\rm Im}}

\def\forE{\mathbb E}
\def\forK{\mathbb K}
\def\capB{\mbox{\boldmath${\mathsf B}$}}
\def\capC{\mbox{\boldmath${\mathsf C}$}}
\def\capD{\mbox{\boldmath${\mathsf D}$}}
\def\capE{\mbox{\boldmath${\mathsf E}$}}
\def\capG{\mbox{\boldmath${\mathsf G}$}}
\def\tcapG{\tilde{\capG}}
\def\capH{\mbox{\boldmath${\mathsf H}$}}
\def\capK{\mbox{\boldmath${\mathsf K}$}}
\def\capL{\mbox{\boldmath${\mathsf L}$}}
\def\capM{\mbox{\boldmath${\mathsf M}$}}
\def\capR{\mbox{\boldmath${\mathsf R}$}}
\def\capW{\mbox{\boldmath${\mathsf W}$}}

%imaginary unit
\def\i{\mbox{${\mathrm i}$}}

\def\mC{\mbox{\boldmath${\mathcal C}$}}

\def\mB{\mbox{${\mathcal B}$}}
\def\mE{\mbox{${\mathcal{E}}$}}
\def\mL{\mbox{${\mathcal{L}}$}}
\def\mK{\mbox{${\mathcal{K}}$}}
\def\mV{\mbox{${\mathcal{V}}$}}

\def\C{\mbox{\boldmath${\mathcal C}$}}
\def\E{\mbox{\boldmath${\mathcal E}$}}

%Definitions: journals
\def\ARMA{{ Arch. Rat. Mech. Analysis\ }}
\def\AMR{{ Appl. Mech. Rev.\ }}
\def\ASCEEM{{ ASCE J. Eng. Mech.\ }}
\def\acta{{ Acta Mater. \ }}
\def\CMAME {{ Comput. Meth. Appl. Mech. Engrg.\ }}
\def\CRAS{{ C. R. Acad. Sci., Paris\ }}
\def\EFM{{ Eng. Fracture Mechanics\ }}
\def\EJMA{{ Eur.~J.~Mechanics-A/Solids\ }}
\def\IJES{{ Int. J. Eng. Sci.\ }}
\def\IJF{{ Int. J. Frac.\ }}
\def\IJMS{{ Int. J. Mech. Sci.\ }}
\def\IJNME{{ Int. J. Num. Meth. Engng.\ }}
\def\IJNAMG{{ Int. J. Numer. Anal. Meth. Geomech.\ }}
\def\IJP{{ Int. J. Plasticity\ }}
\def\IJSS{{ Int. J. Solids Structures\ }}
\def\IngA{{ Ing. Archiv\ }}
\def\JAM{{ J. Appl. Mech.\ }}
\def\JAP{{ J. Appl. Phys.\ }}
\def\JE{{ J. Elasticity\ }}
\def\JM{{ J. de M\'ecanique\ }}
\def\JMM{{ J. Micromech. Microeng.\ }}
\def\JMPS{{ J. Mech. Phys. Solids\ }}
\def\JOMMS{{ J. Mech. Materials Struct.\ }}
\def\Macro{{ Macromolecules\ }}
\def\MOM{{ Mech. Materials\ }}
\def\MMS{{ Math. Mech. Solids\ }}
\def\MPCPS{{ Math. Proc. Camb. Phil. Soc.\ }}
\def\MRC{{ Mech. Res. Comm.\ }}
\def\MSE{{ Mater. Sci. Eng.}}
\def\nature{{ Nature\ }}
\def\PM{{Phil. Mag.\ }}
\def\PMPS{{ Proc. Math. Phys. Soc.\ }}
\def\PRSA{{ Proc. R. Soc. A\ }}
\def\PRSL{{ Proc. R. Soc.\ }}
\def\rock{{ Rock Mech. and Rock Eng.\ }}
\def\QAM{{ Quart. Appl. Math.\ }}
\def\QJMAM{{ Quart. J. Mech. Appl. Math.\ }}

% segue comando pazzesco di zaccaria

\def\salto#1#2{
%\left[\mbox{\hspace{-#1em}}\left[#2\right]\mbox{\hspace{-#1em}}\right]}
[\mbox{\hspace{-#1em}}[#2]\mbox{\hspace{-#1em}}]}

%\def\salto{
%\left[\mbox{\hspace{-#1em}}\left[#2\right]\mbox{\hspace{-#1em}}\right]}
%[[#1]]}

%\newcommand{\salto}[1]{\ds{\|#1\|}}

%dopodiche' scrivi il comando
%\salto{*}{**}
%dove * e' un numero mentre ** e' quello che tu vuoi tra parentesi.
%Il numero serve a far si che le due parentesi quadre che comporranno
%l'unica parentesi che tu vuoi siano ben posizionate. Tale numero varia a
%seconda di chi sia **.

\title{The dynamics of a shear band}

\author{Diana Giarola$^{1}$, Domenico Capuani$^2$, Davide Bigoni$^{1,*}$\\ 
(1) DICAM, University of Trento, via Mesiano 77, I-38050 Trento, Italy\\
(2) DA, University of Ferrara, via Quartieri 8, I-44121 Ferrara, Italy \\
e-mail: diana.giarola@unitn.it, bigoni@ing.unitn.it, domenico.capuani@unife.it\\
}
\date{}
\maketitle
\footnotetext[1]{Corresponding author: Davide Bigoni 
fax: +39 0461 282599; tel.: +39 0461 282507; web-site:
http://www.ing.unitn.it/$\sim$bigoni/; e-mail:
bigoni@ing.unitn.it.}

\begin{abstract} 

A shear band of finite length, formed inside a ductile material at a certain stage of a continued homogeneous strain, provides a dynamic perturbation to an incident
wave field, which strongly influences the dynamics of the material and affects its path to failure. 
The investigation of this perturbation is presented for a ductile metal, with reference to the incremental mechanics of a material obeying the J$_2$--deformation theory of plasticity (a special form of prestressed, elastic, anisotropic, and incompressible solid). 
The treatment originates from the derivation of integral representations relating the incremental mechanical fields at every point of the medium to the incremental displacement jump across the shear band faces, generated by an impinging wave. The boundary integral equations (under the plane strain assumption) are numerically approached through a  collocation technique, which keeps into account the singularity at the shear band tips and permits the analysis of an incident wave impinging a shear band. It is shown that the presence of the shear band induces a resonance, visible in the incremental displacement field and in the stress intensity factor at the shear band tips, 
which promotes shear band growth. Moreover, the waves scattered by the shear band are shown to generate 
a fine texture of vibrations, 
parallel to the shear band line and propagating at a long distance from it, but 
leaving a sort of conical shadow zone, which emanates from the tips of the shear band.

\end{abstract}

\section{Introduction}

When a ductile material is subject to severe strain, failure is preluded by the emergence of shear bands which initially nucleate in a small area, but quickly extend rectilinearly and accumulate damage, until they degenerate into fractures. Therefore, research on shear bands yields a fundamental understanding of the intimate rules of failure\footnote{The fact that shear bands initiate the complex mechanism of failure in ductile materials is the reason for which a great research effort has been devoted to this topic, see among others \cite{merodio, morin, petryk, teko}}, so that it may be important in the design of new materials with superior mechanical performances. 

Modelling of a shear band as a slip plane embedded in a highly prestressed material and perturbed by a mode II incremental strain, reveals that a highly inhomogeneous and strongly focussed stress state is created in the proximity of the shear band and aligned parallel to it. 
This evidence, together with the fact that the incremental energy release rate blows up when the stress state approaches the condition for ellipticity loss, may explain the rectilinear growth of shear bands 
(documented in several experiments, \cite{Guduru, kalthoff, xu, Yang}) 
and the reason why they are a preferred mode of failure for ductile materials 
\cite{bigonifdc, bordy, palmer, puzrin, rice73}.

Although dynamic effects play an important role on shear band growth and related failure development, most of the analyses conducted so far were limited to quasi-static conditions, while numerical simulations addressing the dynamics of shear bands are scarce (and referred to high strain-rate loading \cite{Bonnet, doli, kud, li, lis, nee, Medyanik, vaz, Zhou}). 
The aim of the present article is to investigate dynamic perturbations in the stress/deformation fields of an incident wave, induced by a shear band of finite length, formed inside a ductile metal, at a certain stage of a continued strain. The shear band is modelled as possessing  null thickness and thus behaving as a discontinuity surface, an assumption which is motivated by the experimental observation 
\cite{Qu,Song,Zhang} that thicknesses in metals are on the order of micrometres, while lengths can reach millimeters, 
so that a thickness-to-length ratio of order $10^{-3}$ is considered to be negligible. Hence,
the incremental behaviour of a prestressed, elastic, anisotropic and incompressible material,  containing a finite-length shear band of negligible thickness, is analyzed in the dynamic regime. To this purpose, integral representations are derived (under the plane strain condition\footnote{
Results presented in this article can be extended to three dimensional deformation by using the Green's function derived in \cite{cono}.}
and assuming ellipticity and homogeneity of the material properties), relating the incremental fields at every point of the medium to the incremental displacement jump across the shear band faces, which originates from an impinging wave. 
The integral equations are numerically solved with an {\it ad hoc} developed collocation method, which allows for the treatment of the singularities present at the tips of the shear bands and provides a basis for the analysis of dynamic disturbances propagating in a solid near the boundary of ellipticity loss. The collocation technique was previously used for crack problems \cite{brunbigoni,CuapianiWillis,dominguez,manolis,salvadori, salvadorygray, tan}, not for shear bands, and presents 
several advantages when compared with finite element methods (see for instance \cite{Bonnet, lis, Medyanik}), the main ones being: (i.) only the boundary of the 
shear band is discretized; (ii.) an infinite medium is naturally embodied; (iii.) the radiation damping condition is automatically satisfied, avoiding spurious wave reflection at ficticious boundaries.

Although the developed formulation is general, so that it includes as special cases neo-Hookean, Mooney-Rivlin and Ogden incompressible elasticity \cite{bigoni_libro, biot}, results are presented for a model of ductile metals, namely, the J$_2$--deformation theory of plasticity \cite{hutch}. 
It is shown that when a metal is severely strained (near the border of failure of ellipticity) and a shear band is already formed inside of it (providing a weak surface  
of low stiffness), a complex interaction with incoming waves is developed, showing a narrow focusing of the deformation, which emanates from the shear band tips, and the 
emergence of a fine texture of short wavelength vibrations parallel to the shear band. 
The wave texture extends to a large distance from the shear band and the wavelength vanishes, 
when failure of ellipticity is approached, so that a 
conical shadow zone is evidenced, originating from the tip of the shear band, where the scattered wave field is strongly attenuated. 
Moreover, a resonance wavelength is observed, amplifying the displacement jump across the shear band and the stress intensity factor at its tips, thus promoting shear band growth.

\section{Incremental constitutive equations and dynamic Green's function} \lb{costitutu}

Biot \cite{biot} has shown that the most general incremental response of a hyperelastic incompressible material deformed in plane strain
can be characterized in terms of the Zaremba-Jaumann derivative of the Cauchy stress $\stackrel{\nabla}{\bsigma}$, 
\beq
\lb{biot}
\stackrel{\nabla}{\bsigma} = \dot{\bsigma} -\bW\bsigma+\bsigma \bW,
\eeq
(where $\bW$ is the incremental rotation tensor and $\dot{\bsigma}$ is the increment of Cauchy stress) as
\beq
\stackrel{\nabla}{\sigma}_{11}-\stackrel{\nabla}{\mbox{$\sigma$}}_{22} = 2 \mu_* \left(D_{11}-D_{22}\right), 
~~~~
\stackrel{\nabla}{\mbox{$\sigma$}}_{12} = 2 \mu D_{12}, 
\eeq
where, in an updated Lagrangean description (in which the current state is used as reference), 
$\mu$ and $\mu_*$ are two incremental shear 
moduli, respectively parallel and inclined at 45$^\circ$ with respect to the $x_1-$axis \cite{bigoni_libro}, and $D_{ij}$ are the in-plane components 
of the Eulerian strain increment tensor $\bD$, which has to satisfy the incompressibility constraint $\tr \bD=0$.

In terms of the unsymmetric nominal stress increment $\dot{\bt}$, related to the Zaremba-Jaumann derivative of the Cauchy stress as 
\beq 
\lb{Sdot3_bis}
\dot\bt=\stackrel{\nabla}{\bsigma}-\bsigma\bW-\bD\bsigma, 
\eeq
the constitutive equations (\ref{biot}), together with the incompressibility constraint, can be rewritten as 
\beq 
\lb{traction} 
\dot t_{ij}=\forK_{ijkl}v_{l,k}+\dot p\delta
_{ij},~~~v_{i,i}=0,
\eeq
where, $v_i$ is the incremental displacement, $\dot{p}$ is the incremental hydrostatic stress and $\delta _{ij}$ is the Kronecker delta (indices range between 1 and 2, a comma denotes partial differentiation). The fourth-order tensor
$\forK_{ijkl}$ of the instantaneous moduli, posseses the major symmetry 
$\forK_{ijkl} = \forK_{klij}$ (but not the two minor), and is defined in components as 
\beq 
\label{piola1} 
\barr{lll} 
\forK_{1111} =\mu_{*}-\ds{\frac \sigma 2}-p, & \forK_{1122}=-\mu _{*}, &
\forK_{1112}=\forK_{1121}=0,  \\ [3 mm] 
\forK_{2211} = -\mu_{*} , & \forK_{2222}=\mu
_{*}+\ds{\frac \sigma 2}-p, & \forK_{2212}=\forK_{2221}=0, \\ [3 mm] 
\forK_{1212} = \mu +\ds{\frac \sigma 2}, & \forK_{1221}= \forK_{2112}=\mu -p, & \forK_{2121}=\mu -\ds{\frac\sigma 2} ,  
\earr 
\eeq 
where the prestress parameters $\sigma$ and $p$ are the in-plane deviatoric and mean stresses, functions of the principal Cauchy stresses, respectively, as
\begin{equation}
\label{eq3}
\sigma =\sigma _1-\sigma _2,\qquad p=\frac{\sigma _1+\sigma _2}2.
\end{equation}

The constitutive equations (\ref{traction})--(\ref{piola1}) are representative of a broad class of material behaviours, including all possible elastic incompressible materials which are orthotropic with respect to the current principal stress directions. Note that initial orthotropy may find interesting applications in the field of fibre-reinforced elastic materials.

In this article attention is focused on the behaviour of ductile metals, which can be represented through the J$_2$--deformation theory of plasticity, whose constitutive equations (Hutchinson and Neale, 1979) in plane strain reduce to 
\beq
\lb{tortellini}
\sigma_1-\sigma_2 = K \left(\frac{2}{\sqrt{3}}\right)^{N+1} \, |\varepsilon_1|^{N-1} \varepsilon_1 ,
\eeq
where $K$ is a stiffness parameter, $N \in (0,1]$ a hardening exponent and $\varepsilon_1 = - \varepsilon_2$ are the logarithmic strains, related to the principal stretches $\lambda_1=1/\lambda_2$ via 
$\varepsilon_1 = \log \lambda_1 = - \varepsilon_2 = -\log \lambda_2$. 
The incremental moduli $\mu$ and $\mu_*$, defining equation (\ref{traction}), can be written as
\beq
\lb{modulazzicertamente}
\mu  = \ds{\frac{1}{3}E_s\left(\varepsilon_1 - \varepsilon_2\right)
\coth\left(\varepsilon_1 - \varepsilon_2\right)}, ~~~~~
\mu_*=\ds{\frac{1}{9}\frac{E_s}{\varepsilon_e^2}\left[3{\left(\varepsilon_1+\varepsilon_2\right)}^2+
N{\left(\varepsilon_1-\varepsilon_2\right)}^2\right]},
\eeq
where $E_s$ is the secant modulus to the effective-stress/effective-strain curve, given by 
\beq
E_s = K \left(\frac{2}{\sqrt{3}}\right)^{N-1}|\varepsilon_1|^{N-1}.
\eeq

Note that the out-of-plane stress increment can be calculated from the expression
\beq
\stackrel{\nabla}{\mbox{$\sigma$}}_{33} = \tr  \dot{\bsigma}/3 = \dot{p}.
\eeq

The equations of incremental motion can be written as 
\beq 
\lb{eqmot} 
\dot{t}_{ij,i}+\dot{f}_j=\rho \frac{\partial^2 v_j}{\partial t^2}, 
\eeq
where $\rho$ is the mass density, $\dot{f}_j$ the incremental body force, and $t$ denotes the time. 
For time-harmonic motion with circular frequency $\Omega$, the incremental displacement field $v_i (\bx) \exp(-i \Omega t)$ can be derived from a stream function $\psi(\bx)\exp(-i \Omega t)$, introduced as
\beq 
\lb{stream} 
v_1=\psi_{,2},\quad v_2=-\psi_{,1}.
\eeq
A substitution of equation (\ref{traction}) in equation (\ref{eqmot}) yields the differential equation 
\begin{multline} 
\lb{eqdiff}
 \left(1+k\right)\psi_{,1111}+2\left(2\xi-1\right)\psi_{,1122}+\left(1-k\right)\psi_{,2222}+\frac{\dot{f}_{1,2}}{\mu}-\frac{\dot{f}_{2,1}}{\mu}+\\+\frac{\rho}{\mu} \Omega^2\left(\psi_{,11}+\psi_{,22}\right) = 0, 
\end{multline}
where $k$ is a dimensionless measure of the deviatoric pre-stress and $\xi$ a dimensionless parameter quantifying the amount of orthotropy 
\beq 
k=\frac{\sigma}{2 \mu}, \qquad \xi=\frac{\mu_*}{\mu},
\eeq
which for the J$_2$--deformation theory of plasticity (\ref{modulazzicertamente}) become
\beq
k = \frac{1}{\coth(2\varepsilon_1)}, ~~~~~ \xi = \frac{N}{2\varepsilon_1\coth(2\varepsilon_1)}.
\eeq

The principal part of the differential equation (\ref{eqdiff}) is the same of the quasi-static case and defines the regime classification in terms of the following coefficients
\beq 
\left.
\begin{array}{l}
\gamma_1 \\ [3 mm]
\gamma_2
\end{array}
\right\} =
\frac{1-2 \mu_*/\mu\pm\sqrt{\left(1-2 \mu_*/\mu\right)^2 + k^2-1}}{1+k} ,
\eeq
so that either two real and negative (in the so-called \lq elliptic imaginary regime' denoted by EI) or two complex-conjugate 
(in the so-called \lq elliptic complex regime' denoted by EC) 
coefficients are only possible in the elliptic range, to which the present study is restricted. 

Note that for a tensile prestress, the Hill condition, which excludes every incremental bifurcation \cite{bigoni_libro, hill,radibigcapu}, is given by (for $\mu>0$)
\beq 
\lb{excl}
0< p/\mu < 2 \xi,\qquad \frac{k^2+(p/\mu)^2}{2 p/\mu}< 1, 
\eeq
while a surface instability occurs for compressive prestress when \cite{bigoni_libro, hillhu}
\beq 
4 \xi - 2 p/\mu=\frac{(p/\mu)^2 -2 p/\mu +k^2}{\sqrt{1-k^2}}.
\eeq

The infinite body Green's functions can be found by solving equation (\ref{eqdiff}) when the body force is given by the Dirac delta function $\delta(\bx)$, i.e. $\dot{f}_j \delta(\bx)$. 
Introducing a plane wave expansion for the incremental displacement $v_i^g$ of the Green's state 
\beq 
\lb{incrdisp}
v_i^g(\bx)=-\frac{1}{4\pi^2}\oint_{|\bomega
=1|}\tilde{v}_i^g(\bomega \scalp\bx)d\bomega,
\eeq 
the following representation can be obtained from equation (\ref{eqdiff}) in the transformed domain 
\beq 
\label{vtrans}
\barr{cl} 
\tilde v_i^g(\bomega\scalp\bx) = & \ds{\frac {( \delta_{1i}\omega_2
-\delta_{2i}\omega_1)( \delta_{1g}\omega_2
-\delta_{2g}\omega_1)}{L(\bomega)}
}\hspace{1mm}[\mbox{Ci}(\eta\mid\bomega\scalp\bx\mid)\cos{(\eta\hspace{1mm}\bomega\scalp\bx)}+\\[5mm]
&+
\mbox{Si}(\eta\hspace{1mm}\bomega\scalp\bx)\sin{(\eta\hspace{1mm}\bomega\scalp\bx)}-i\ds{\frac{\pi}{2}
}\cos{(\eta\hspace{1mm}\bomega\scalp\bx)}], 
\earr 
\eeq 
where Ci and Si are the cosine integral and sine integral functions, respectively, and 
\beq 
L(\bomega)=\mu(1+k)\omega_2^4\left(\frac{\omega_1^2}{\omega_2^2}-\gamma_1\right)\left(\frac{\omega_1^2}{\omega_2^2}-\gamma_2\right)>0,
\eeq 
with 
\beq \eta=\Omega\sqrt{\frac{\rho}{L(\bomega)}}.
\eeq

The gradient of the incremental displacement (\ref{incrdisp}) can be written as 
\beq 
\lb{gradincrdisp} 
v_{i,k}^g(\bx)=-\frac{1}{4\pi^2}\oint_{|\bomega|=1}\tilde{v}_{i,k}^g(\bomega \scalp\bx)d\omega 
\eeq
where 
\beq 
\tilde{v}_{i,k}^g(\bomega \scalp\bx)=\omega_k\frac{\delta_{ig}-\omega_i\omega_g}{L(\bomega)}\left[\frac{1}{\bomega \scalp \bx}-\eta \Xi(\eta \bomega \scalp \bx)\right]
\eeq
and
\beq
\Xi(\alpha)=\sin(\alpha) \mbox{Ci}(|\alpha|)-\cos(\alpha) \mbox{Si}(\alpha)-i \frac{\pi}{2}\sin(\alpha).
\eeq
The plane wave expansion (\ref{eqdiff}) has been developed in \cite{bigonicapuani02} and \cite{bigonicapuani05}.
Finally the Green's function for incremental nominal stresses can be derived from the constitutive equations (\ref{traction}) as 
\beq 
\lb{green_ns} 
\barr {ll}
\dot{t}_{11}^g = \left(2\mu_*-p\right)v^g_{1,1}+\dot\pi^g,\,\,\, &
\dot{t}_{12}^g = \left(\mu-p\right)v^g_{1,2}+\left(\mu+\mu k\right)v^g_{2,1}, \\[2 mm]
\dot{t}_{21}^g = \left(\mu-p\right)v^g_{2,1}+\left(\mu-\mu
k\right)v^g_{1,2},\,\,\,  &
\dot{t}_{22}^g = -\left(2\mu_*-p\right)v^g_{1,1}+\dot\pi^g. 
\earr 
\eeq

\section{The shear band model}

\begin{figure}[!h]
  \begin{center}
    \includegraphics[width=1\textwidth]{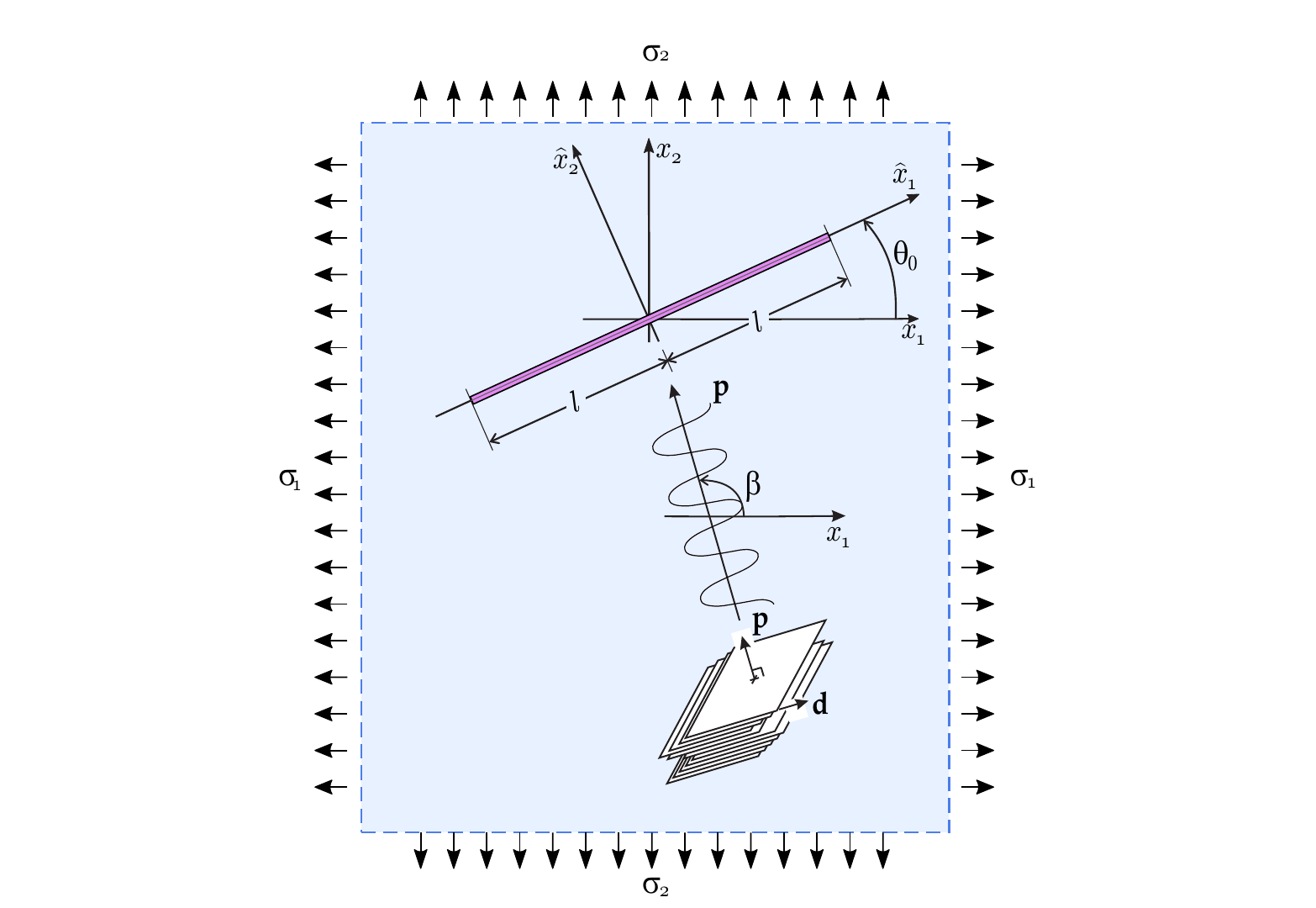}
    \caption{\footnotesize A plane shear wave (sketched as a moving deck of cards) is impinging a shear band of finite-length ($2l$) in a prestressed, orthotropic material. The shear band (aligned parallel to the $\hat{x}_1$--axis) is inclined at an angle $\vartheta$ (positive when anticlockwise) with respect to the orthotropy axes $x_1$ and $x_2$ (aligned parallel to the prestress state); the wave is inclined at an angle $\beta$ with respect to the $x_1$-axis.}
    \label{inclinatosb}
  \end{center}
\end{figure}

A shear band of finite length $2l$ is a very thin layer of material subject to intense shear, emerging inside a ductile material at a certain stage of a uniform deformation path with a well-defined inclination, measured from the $\sigma_1$--principal axis of stress in terms of the angle $\theta_0$ defined as
\beq 
\theta_0 = \mbox{arccot} \sqrt{\frac{1+2\,\text{sign}(k) \sqrt{\xi(1-\xi)}}{1-2 \xi}},
\eeq
and calculated at ellipticity loss \cite{bigoni_libro, hillhu} within the constitutive framework presented in Section \ref{costitutu}.
The shear band is characterized by a high compliance to shear parallel to it, so that it can be modelled by assuming that the incremental nominal traction tangential to the shear band vanishes,
while the normal nominal traction and the normal component of the incremental displacement remain continuous. 
Introducing 
the jump operator $\salto{0.05}{~}$ as
\beq
\salto{0.05}{g} = g^+-g^-,
\eeq
[where $g^+$ and $g^-$ denote the limits approached by the field $g(\bx)$ at the discontinuity surface] 
and two reference systems, namely, 
$x_1-x_2$ aligned parallel to the orthotropy axes of the material and $\hat{x}_1-\hat{x}_2$ aligned parallel to the shear band, Fig. \ref{inclinatosb}, the conditions 
holding along the shear band are the following.
\begin{itemize}
\item Null incremental nominal shearing tractions:
\beq
\lb{condizioni1}
\hat{t}_{21}(\hat{x}_1,0^\pm)=0, ~~~ \forall |\hat{x}_1|< l.
\eeq
\item Continuity of the incremental nominal traction orthogonal to the shear band:
\beq
\lb{condizioni2}
\salto{.05}{\hat{t}_{22}(\hat{x}_1,0)}= 0, ~~~  \forall |\hat{x}_1|< l.
\eeq
\item Continuity of the incremental displacement component orthogonal to the shear band:
\beq
\lb{condizioni3}
\salto{.05}{\hat{v}_{2}(\hat{x}_1,0)} = 0, ~~~  \forall |\hat{x}_1|< l.
\eeq
\end{itemize}

The above equations show that the shear band is modelled as a (null-thickness) discontinuity surface, which is more general than a crack (because a shear band can carry a finite compressive tractions across his faces), but may represent a dislocation \cite{argani, Xiaohan1, Xiaohan2}; in metals the null-thickness assumption is 
strongly motivated by the experimental observation 
\cite{Qu, Song, Zhang} that a shear band thickness-to-length ratio is of the order  $10^{-3}$ since 
lengths of shear bands can reach millimeters, while their thickness is confined to only a few micrometres.
In the absence of prestress, the shear band model reduces to a weak surface whose faces can freely slide and at the same time are constrained to remain in contact, but when a prestress is present, the shear band model differs from that of a sliding planar surface \cite{bordi}. The prescriptions (\ref{condizioni1})--(\ref{condizioni3}) have been directly borrowed from those defining the onset of a shear band in a material \cite{bigoni_libro}.

%%%%%%%%%%%%%%%%%%%%%%%%%%%%%%%%%%%%%%%%%%%%%%%%%%%%%%%%%%%%%%%%%
%%%%%%%%%%%%%%%%%%%%%%%%%%%%%%%%%%%%%%%%%%%%%%%%%%%%%%%%%%%%%%%%%

The analysis of the shear band will be restricted to a tensile prestress and the value of the normalized in-plane mean stress,  
$
p/\mu, 
$
will be selected 
in such a way that the Hill exclusion condition (\ref{excl}) is satisfied, so that spurious bifurcations will not affect the 
dynamics near the shear band, which is analyzed as follows.

Assuming a time-harmonic motion of circular frequency $\Omega$, a wave characterized by an incremental displacement field $\bv^{inc}(\bx) e^{-i\Omega t}$
travels through the medium and is incident upon the shear band. Then, a scattered incremental displacement field $\bv^{sc}(\bx) e^{-i\Omega t}$ 
is generated by the interaction of the incident wave with the shear band such that the total incremental displacement field $\bv(\bx) e^{-i\Omega t}$ is represented as the sum
\beq
\bv = \bv^{inc} + \bv^{sc}.
\eeq

The scattered field $\bv^{sc}$ must satisfy the radiation condition at infinity and the conditions of energy boundedness near the shear band edge. Outside the shear band, the incremental displacement field satisfies the equations of motion, which written in terms of the stream function $\psi$ reduce to equation (\ref{eqdiff}) with null body force. 

The incident wave field is represented by an incremental, time-harmonic plane wave propagating with phase velocity $c$ in a direction defined by the unit propagation vector $\bp$ \cite{ogden} and having
the form
\beq
\lb{ondazza}
\bv^{inc} = A\bd e^{i\frac{\Omega}{c}(\bx\scalp \bp-c t)},
\eeq
where $A$ is the amplitude, $\bd$ is the direction of motion and  $c$ the wave velocity. 
Since the wave (\ref{ondazza}) propagates in an incompressible material, isochoricity implies 
\beq
\lb{oooohhh}
\bd\scalp\bp = 0,
\eeq
so that the incident wave is transverse, with the motion orthogonal to the propagation direction. 
A substitution of equation (\ref{ondazza}) into equation (\ref{eqdiff}), written with $\dot{f}_{1,2}=\dot{f}_{2,1}=0$, and use of equation (\ref{oooohhh}) yield the following expression for the wave speed
\beq
c^2 = \frac{\mu}{\rho} \left[ (1+k) p_1^4+ 2 \left(2 \xi - 1 \right) p_1^2p_2^2 + (1-k) p_2^4 \right],
\eeq
which, setting $p_1=\cos\beta$ and $p_2=\sin\beta$ and 
\beq
c_1= \sqrt{\mu(1+k)/\rho},
\eeq
provides
\beq\label{calfa}
c(\beta) = c_1 \sin^2\beta \sqrt{\left(\cot^2\beta - \gamma_1 \right) \left(\cot^2\beta - \gamma_2 \right) }.
\eeq
Note that in the limits $\beta \rightarrow 0$ and $\beta \rightarrow \pi$, $c$ tends to $c_1$, which represents the speed of a wave traveling in the direction of the $x_1$--axis.

\section{Integral representation for the dynamics of a shear band}

The scattered field $\bv^{sc}$ satisfies the extension of the Betti identity provided in \cite{bigonicapuani05, stokes} 
\beq
\lb{betti}
v_g^{sc}(\by) = \int_{\partial B} \left(\dot{t}_{ij}n_i v_j^g(\bx, \by) - \dot{t}^g_{ij}(\bx, \by) n_i v_j \right) dl_{\bx} ,
\eeq
where $\partial B$ represents the boundary of the shear band, which is made up of two straight lines of length $2l$, with external unit normals of opposite sign, so that equation (\ref{betti}) 
can be specialized for a shear band to  
\beq
\lb{scamorza}
v_g^{sc}(\by) = - \int_{-l}^l \left(\salto{0.38}{\dot{t}_{ij}}n_i v_j^g(\hat{x}_1, \by) - \dot{t}^g_{ij}(\hat{x}_1, \by) n_i \salto{0.38}{v_j} \right) d\hat{x}_1.
\eeq
Because the incremental traction is continuous across the shear band, equations (\ref{condizioni1})--(\ref{condizioni2}), the following boundary integral equation
is obtained
\beq
\lb{scamorza2}
v_g^{sc}(\by) = \int_{-l}^l \dot{t}^g_{ij}(\hat{x}_1, \by) n_i \salto{0.38}{v_j} \, d\hat{x}_1,
\eeq
which provides the incremental displacement at every point in the body as function of the jump of the incremental displacement $\salto{0.38}{v_j}$ across the shear band. 

The gradient of the incremental displacement can be evaluated from the integral equation (\ref{scamorza2}) as
\beq
\lb{scamorza3}
v_{g,k}^{sc}(\by) = -\int_{-l}^l \dot{t}^g_{ij,k}(\hat{x}_1, \by) n_i \salto{0.38}{v_j} \, d\hat{x}_1,
\eeq
so that from the constitutive equations (\ref{traction}) the incremental stress can be written as 
\beq
\dot{t}_{lm}^{sc}(\by) = -\mathbb{K}_{lmkg}  \int_{-l}^l \dot{t}^g_{ij,k}(\hat{x}_1, \by) n_i \salto{0.38}{v_j} \, d\hat{x}_1 +\dot{p}(\by) \delta_{lm},
\eeq
where the incremental in-plane mean stress $\dot{p}$, for the moment unknown, can be determined from the following boundary integral equation \cite{bigonicapuani07}
\beq
\lb{peretta}
\begin{array}{rcl}
	\dot{p}(\by) &=& \ds{ -\int_{\partial
			B}\dot{t}_{ig}\,n_i\,\dot{p}^g(\bx-\by)\,dl_x
		+ \int_{\partial B}n_i\,v_j\,\forK_{ijkg}\,\dot{p}^g_{,k}(\bx-\by)\,
		dl_x +}\\ [5 mm] 
		&&-\ds{\int_{\partial B}\,v_i\,n_i\left[\left(4\mu\mu_*-4\mu^2_*+\mu
		\sigma-2\mu_*\sigma -\frac{\sigma^2}{2}\right) \right.
		v^1_{1,11}(\bx-\by)+}\\[5 mm]
		&&\ds{\left. -\,\sigma\left(\mu+\frac{\sigma}{2}\right) \,v^2_{2,11} (\bx-\by)+\rho \Omega^2 W(\bx-\by)\right] dl_x },
\end{array}
\eeq
where $\dot{p}^g$ is the incremental in-plane mean stress of the Green's state 
\beq
\dot{p}^g =\dot{\pi}^g-\frac{\sigma}{2}v^g_{1,1},
\eeq
in which 
\begin{multline}
\dot{\pi}^g=\dfrac{\omega_g(2 \mu_*-\mu)(1-\omega_g^2)+(\mu-(\delta_{2g}-\delta_{1g})\frac{\sigma}{2})\omega_g^2}{L(\bomega)} 
\left[\dfrac{1}{\bomega \scalp \bx}-\eta \Xi(\eta \bomega \scalp \bx)\right]+\\\vspace{5mm} +\omega_g \eta \Xi(\eta\, \bomega\scalp \bx)
\end{multline}
and 
\beq
\tilde{W}=\left[4\left(\mu-\mu_*\right)\omega^2_2-\sigma\right]\tilde{v}^2_2(\bomega\cdot\bx)+\log |\bomega\cdot\bx| .
\eeq
Introducing for $\partial B$ the straight boundary of the shear band, $\hat{x}_1 \in [-l,l]$, equation (\ref{peretta}) becomes 
\beq
\begin{array}{rcl}
\dot{p}(\by) &=& \ds{ -\int_{\partial
		B}\salto{0.05}{ \dot{t}_{ig}}\,n_i\,\dot{p}^g(\bx-\by)\,dl_x
	+ \int_{\partial B}n_i\,\salto{0.05}{ v_j}\,\forK_{ijkg}\,\dot{p}^g_{,k}(\bx-\by)\,
	dl_x +}\\ [5 mm] 
&&-\ds{\int_{\partial B}\,\salto{0.05}{ v_i}\,n_i\left[\left(4\mu\mu_*-4\mu^2_*+\mu
	\sigma-2\mu_*\sigma -\frac{\sigma^2}{2}\right) \right.
	v^1_{1,11}(\bx-\by)+}\\[5 mm]
&&\ds{ -\,\sigma \left. \left( \mu+\frac{\sigma}{2} \right) \,v^2_{2,11} (\bx-\by)+\rho \Omega^2 W(\bx-\by) \right] dl_x },
\end{array}
\eeq
which, considering the continuity of incremental tractions, equations (\ref{condizioni1})--(\ref{condizioni2}), and the continuity of the normal component of the incremental displacement across the shear band (\ref{condizioni3}) reduces to 
\beq
\dot{p}(\by) = \int_{-l}^l n_i\,\salto{0.05}{ v_j}\,\forK_{ijkg}\,\dot{p}^g_{,k}(\hat{x}_1, \by)\, d\hat{x}_1. 
\eeq

In order to determine the incremental displacement jump $\salto{0.38}{v_j}$, unknown in equation (\ref{scamorza2}), the point $\by$ is assumed to approach the shear band boundary. 
Denoting with $\bs$ the unit vector tangent to the shear band, the boundary conditions at the shear band become
\beq
\lb{contorno}
\bn\scalp\dot{\bt}^{(sc)}\bs = - \bn\scalp\dot{\bt}^{(inc)}\bs,
\eeq
so that equation (\ref{scamorza2}) can be rewritten as 
\beq
\lb{palle}
\hat{t}^{(inc)}_{21}(\by) = n_l s_m \mathbb{K}_{lmkg}  \int_{-l}^l \dot{t}^g_{ij,k}(\hat{x}_1, \by) n_i \salto{0.38}{v_j} \, d\hat{x}_1.
\eeq
Equation (\ref{palle}) represents the boundary integral formulation for the dynamics of a shear band interacting with an impinging wave. The kernel of the integral equation (\ref{palle}) is hypersingular of order $r^{-2}$ as $r\rightarrow 0$, being $r$ the distance between field point $\bx$ and source point $\by$
\beq
r = |\bx-\by| = \sqrt{(x_1-y_1)^2-(x_2-y_2)^2}. 
\eeq
Note that the integral on right-hand side of equation (\ref{palle}) is specified in the finite-part Hadamard sense.

The solution for an inclined shear band in an infinite medium
can be expressed in the inclined reference system sketched in Fig. \ref{inclinatosb}.

The components of the vector of incremental displacements $\bv$  in the
reference system $x_1$--$x_2$, can be expressed in the local reference system $\hat{x}_1$--$\hat{x}_2$ as
\beq
\lb{giragiragira1}
 \bv = \bQ\hat{\bv}, ~~~ [\bQ] = \left[\barr{cc}\cos \vartheta &  -\sin \vartheta\\   \sin \vartheta & \cos \vartheta \earr\right],
\eeq
so that, due to the boundary conditions (\ref{condizioni3})  
\beq
\salto{0.38}{v_j} = Q_{j1}\salto{0.38}{\hat{v}_1} = s_j \salto{0.38}{\hat{v}_1}, 
\eeq
equation (\ref{palle}) can be given the final form 
\beq
\lb{pallissime}
\hat{t}_{21}^{(inc)}(\by) = n_l s_m \mathbb{K}_{lmkg}  \int_{-l}^l \dot{t}^g_{ij,k}(\hat{x}_1, \by) n_i s_j \, \salto{0.38}{\hat{v}_1} \, d\hat{x}_1,
\eeq
showing that the dynamics of a shear band is governed by a linear integral equation in the unknown jump of tangential incremental displacement across the shear band faces, $\salto{0.38}{\hat{v}_1}$. It is worth noting that the gradient of the Green incremental stress tensor, constituting the kernel of the boundary integral equation, turns out to be the sum of a static part $\dot{t}_{ij,k}^{g (st)}$ and a dynamic part $\dot{t}_{ij,k}^{g (dyn)}$, whose expressions are given in Appendix A, leading to
\beq
\lb{pallissime2}
\hat{t}_{21}^{(inc)}(\by) = n_l s_m \mathbb{K}_{lmkg}  \int_{-l}^l \left(\dot{t}_{ij,k}^{g (st)}(\hat{x}_1, \by)+\dot{t}_{ij,k}^{g (dyn)}(\hat{x}_1, \by)\right) n_i s_j \, \salto{0.38}{\hat{v}_1} \, d\hat{x}_1. 
\eeq

Note that a simplification of the equation (\ref{pallissime}) in the case of a horizontal shear band ($\theta=0$) is provided in Appendix B.

\section{Boundary integral equations and numerical procedure}

The treatment of the boundary integral equation (\ref{pallissime2}) requires the development of an {\it ad hoc} numerical procedure, which needs the implementation of a special strategy 
to enforce the singular behaviour at the band tips, similar to that developed for cracks in \cite{salvadori, salvadorygray}.

Since both field and source points $\bx$ and $\by$ lie on the $\hat{x}_1$--axis, equation (\ref{pallissime}) can be rewritten as
\beq
\lb{pallissime4}
\hat{t}_{21}^{(inc)}(\hat{y}) = n_l s_m \mathbb{K}_{lmkg}  \int_{-l}^l \dot{t}_{ij,k}^g (\hat{x}, \hat{y}) n_i s_j \, \salto{0.38}{\hat{v}}(\hat{x}) \, d\hat{x},
\eeq
where the index \lq 1' has been dropped, so that $\hat{x}$, $\hat{y}$ and $\salto{.1}{\hat{v}}$ replace respectively $\hat{x}_1$ , $\hat{y}_1$ and  $\salto{.1}{\hat{v_1}}$.

\begin{figure}[!h]
	\begin{center}
		\includegraphics[width=1\textwidth]{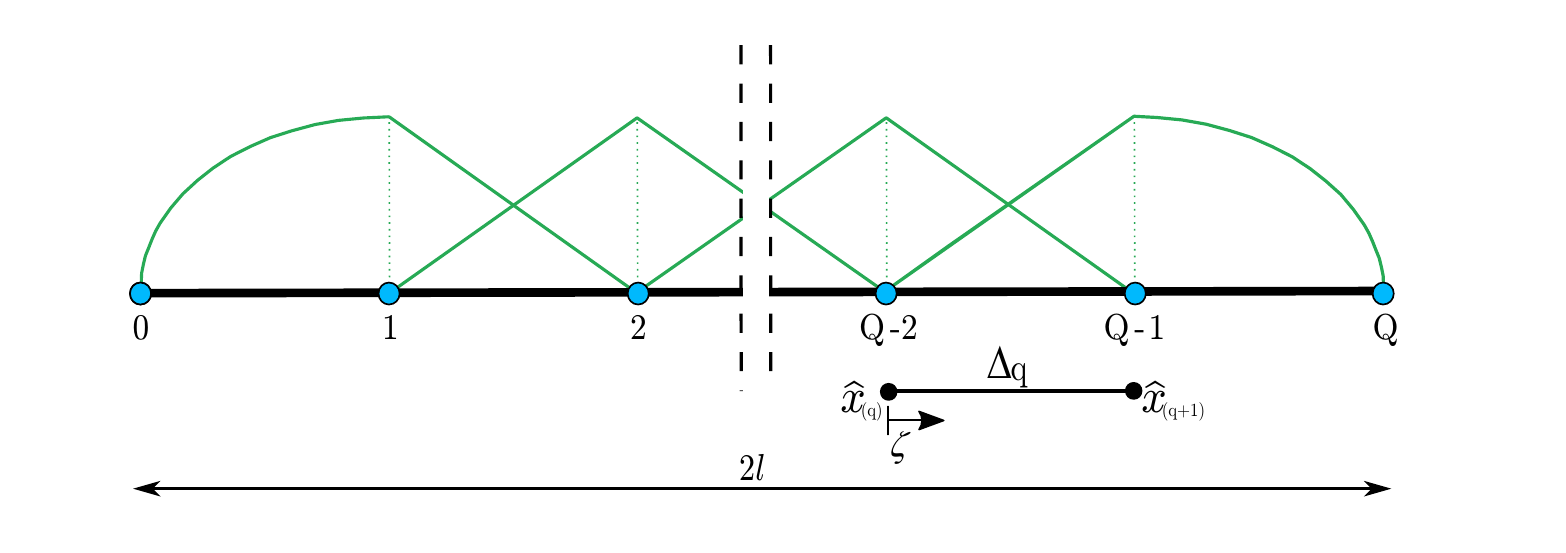}
		\caption{\footnotesize The shear band line is divided in $Q$-intervals. Within each interval a linear variation of the incremental displacement jump is assumed, with the exception of the two intervals at the shear band tips.}
		\label{shape}
	\end{center}
\end{figure}

The shear band segment is divided into Q intervals $[\hat{x}_{(q)},\hat{x}_{(q+1)}]$ $(q=0,\dots,Q-1; \hat{x}_{(0)}=-l,;\hat{x}_{(Q)}=l)$ and a linear variation of the incremental displacement jump $\salto{0.38}{\hat{v}}$ is assumed within each interval, with the exception of the two intervals situated at the shear band tips, where a square root variation of the incremental displacement jump $\salto{0.38}{\hat{v}}$ is adopted:
\beq 
\lb{lin} 
\salto{.1}{\hat{v}}(\hat{x}_{(q)}+\zeta \Delta_q)=\salto{.1}{\hat{v}}_{(q)} (1-\zeta)+\salto{.1}{\hat{v}}_{(q+1)} \zeta \quad (q=1,\dots,Q-2),
\eeq
\beq 
\lb{lin2} 
\salto{.1}{\hat{v}}(\hat{x}_{(q)}+\zeta \Delta_q)=\salto{.1}{\hat{v}}_{(q+1)} \sqrt{\zeta} \quad (q=0),
\eeq
\beq 
\lb{lin3} 
\salto{.1}{\hat{v}}(\hat{x}_{(q)}+\zeta \Delta_q)=\salto{.1}{\hat{v}}_{(q)} \sqrt{1-\zeta} \quad (q=Q-1),
\eeq
where $\Delta_q=|\hat{x}_{(q+1)}-\hat{x}_{(q)}|$, $\zeta \in [0,1]$ and $\salto{.1}{\hat{v}}_{(q)} (q=1,\dots,Q-1)$ is the nodal value of the incremental displacement jump (Figure \ref{shape}).
%%%%%%%%%%%%%%%%%%%%%%%%%%%%%%%%%%%%%%%%%%%%%%%%%%%%%%%%%%%%%%%%%%%%%%%%%%%%%%%%%%%
%%%%%%%%%%%%%%%%%%%%%%%%%%%%%%%%%%%%%%%%%%%%%%%%%%%%%%%%%%%%%%%%%%%%%%%%%%%%%%%%%%%%%
The square root variation is adopted to take into account the singularity at the shear band tip, as is usual for the crack tip problem \cite{paulino,salvadori,salvadorygray,tan}.

When $\hat{y}=\hat{x}_{(p)} \, (p=1,\dots,Q-1)$ is assumed to be the source point, the relevant integral equation becomes
\begin{multline}\lb{y=x p interno}
	\hat{t}_{21}^{(inc)}(\hat{x}_{(p)}) = \displaystyle n_l s_m \mathbb{K}_{lmkg} n_i s_j  \Delta_0 \int_{0}^1 \dot{t}_{ij,k}^g (\hat{x}_{(0)}+\zeta \Delta_0, \hat{x}_{(p)}) \, \salto{0.38}{\hat{v}}_{(1)}\sqrt{\zeta} \, d\zeta+\\
+\displaystyle n_l s_m \mathbb{K}_{lmkg} n_i s_j \sum_{q=1}^{p-2} \Delta_q \int_{0}^1 \dot{t}_{ij,k}^g (\hat{x}_{(q)}+\zeta \Delta_q, \hat{x}_{(p)}) (\, \salto{0.38}{\hat{v}}_{(q)} (1-\zeta)+ \salto{0.38}{\hat{v}}_{(q+1)}\zeta \,) \, d\zeta+\\
+\displaystyle n_l s_m \mathbb{K}_{lmkg} n_i s_j \sum_{q=p-1}^{p} \Delta_q \int_{0}^1 \dot{t}_{ij,k}^g (\hat{x}_{(q)}+\zeta \Delta_q, \hat{x}_{(p)}) (\, \salto{0.38}{\hat{v}}_{(q)} (1-\zeta)+ \salto{0.38}{\hat{v}}_{(q+1)}\zeta \,) \, d\zeta+\\
+\displaystyle n_l s_m \mathbb{K}_{lmkg} n_i s_j \sum_{q=p+1}^{Q-2} \Delta_q \int_{0}^1 \dot{t}_{ij,k}^g (\hat{x}_{(q)}+\zeta \Delta_q, \hat{x}_{(p)}) (\, \salto{0.38}{\hat{v}}_{(q)} (1-\zeta)+ \salto{0.38}{\hat{v}}_{(q+1)}\zeta \,) \, d\zeta+\\
+\displaystyle n_l s_m \mathbb{K}_{lmkg} n_i s_j  \Delta_{Q-1} \int_{0}^1 \dot{t}_{ij,k}^g (\hat{x}_{(Q-1)}+\zeta \Delta_{Q-1}, \hat{x}_{(p)})\, \salto{0.38}{\hat{v}}_{(Q-1)} \sqrt{1-\zeta} \, d\zeta.
\end{multline}
In equation (\ref{y=x p interno}), the integrals which are singular for $\hat{x}_{(q)}+\zeta \Delta_q \rightarrow \hat{x}_{(p)}$ are relevant to the static kernel $\dot{t}_{12}^{g (st)}$ and can be rearranged as
\beq
\begin{array}{c}
\lb{singularintegral}
\displaystyle \sum_{q=p-1}^{p} \Delta_q \int_{0}^1 \dot{t}_{ij,k}^{g (st)} (\hat{x}_{(q)}+\zeta \Delta_q, \hat{x}_{(p)}) (\, \salto{0.38}{\hat{v}}_{(q)} (1-\zeta)+ \salto{0.38}{\hat{v}}_{(q+1)}\zeta \,) \, d\zeta=\\
\displaystyle\Delta_{p-1} \int_{0}^1 \dot{t}_{ij,k}^{g (st)} (\hat{x}_{(p-1)}+\zeta \Delta_{p-1}, \hat{x}_{(p)})  \salto{0.38}{\hat{v}}_{(p)} \zeta \, d\zeta+\\
\displaystyle+\Delta_{p} \int_{0}^1 \dot{t}_{ij,k}^{g (st)} (\hat{x}_{(p)}+\zeta \Delta_{p}, \hat{x}_{(p)})  \salto{0.38}{\hat{v}}_{(p)} (1-\zeta) \, d\zeta+\\
\displaystyle+\Delta_{p-1} \int_{0}^1 \dot{t}_{ij,k}^{g (st)} (\hat{x}_{(p-1)}+\zeta \Delta_{p-1}, \hat{x}_{(p)})  \salto{0.38}{\hat{v}}_{(p-1)} (1-\zeta) \, d\zeta+\\
\displaystyle+\Delta_{p} \int_{0}^1 \dot{t}_{ij,k}^{g (st)} (\hat{x}_{(p)}+\zeta \Delta_{p}, \hat{x}_{(p)})  \salto{0.38}{\hat{v}}_{(p+1)} \zeta \, d\zeta,
\end{array}\eeq
so that, by means of a change of variable, the integrals can be evaluated as
\beq\begin{array}{c}\lb{hada1}
	\displaystyle \sum_{q=p-1}^{p} \Delta_q \int_{0}^1 \dot{t}_{ij,k}^{g (st)} (\hat{x}_{(q)}+\zeta \Delta_q, \hat{x}_{(p)}) (\, \salto{0.38}{\hat{v}}_{(q)} (1-\zeta)+ \salto{0.38}{\hat{v}}_{(q+1)}\zeta \,) \, d\zeta=\\
	\displaystyle \int_{-\Delta_{p-1}}^{\Delta_{p}} \dot{t}_{ij,k}^{g (st)} (r \be_r)\,  \salto{0.38}{\hat{v}}_{(p)}  \, dr- \left(\frac{1}{\Delta_{p-1}}+\frac{1}{\Delta_{p}}\right)\int_{-\Delta_{p-1}}^{\Delta_{p}} \dot{t}_{ij,k}^{g (st)} (r \be_r)\,  \salto{0.38}{\hat{v}}_{(p)} \, r \, dr+\\
	\displaystyle -\frac{1}{\Delta_{p-1}}\int_{-\Delta_{p-1}}^0 \dot{t}_{ij,k}^{g (st)} (r \be_r)  \, \salto{0.38}{\hat{v}}_{(p-1)} \, r \, dr+\frac{1}{\Delta_{p}} \int_{0}^{\Delta_{p}} \dot{t}_{ij,k}^{g (st)} (r \be_r)\,  \salto{0.38}{\hat{v}}_{(p+1)}\, r \, dr,
\end{array}\eeq
with $\be_r=\mathbf{r} / r$. 
The non-null finite parts of the above integrals can be calculated as 
\beq
\begin{array}{rcl}
\lb{hada2}
	\displaystyle \int_{-\Delta_{p-1}}^{\Delta_{p}} \dot{t}_{ij,k}^{g (st)} (r \be_r)\,  \salto{0.38}{\hat{v}}_{(p)}  \, dr&=&\displaystyle
	T^g_{ijk}(\theta)\left(-\frac{1}{\Delta_{p-1}}-\frac{1}{\Delta_{p}}\right)\,  \salto{0.38}{\hat{v}}_{(p)},\\
	\displaystyle -\frac{1}{\Delta_{p-1}}\int_{-\Delta_{p-1}}^0 \dot{t}_{ij,k}^{g (st)} (r \be_r)  \, \salto{0.38}{\hat{v}}_{(p-1)} \, r \, dr&=&\displaystyle
	T^g_{ijk}(\theta)\frac{\log{\Delta_{p-1}}}{\Delta_{p-1}}\,  \salto{0.38}{\hat{v}}_{(p-1)},\\
	\displaystyle \frac{1}{\Delta_{p}} \int_{0}^{\Delta_{p}} \dot{t}_{ij,k}^{g (st)} (r \be_r)\,  \salto{0.38}{\hat{v}}_{(p+1)}\, r \, dr&=&\displaystyle
	T^g_{ijk}(\theta)\frac{\log{\Delta_{p}}}{\Delta_{p}}\,  \salto{0.38}{\hat{v}}_{(p+1)}
\end{array}\eeq
and the function $T^g_{ijk}(\theta)$ is explicitly provided in Appendix A.

In the particular cases when $\hat{y}=\hat{x}_{(p)}$ is assumed to be the source point and $p=1$ or $p=Q-1$, equation (\ref{y=x p interno}) has to be rewritten as
\begin{multline}\lb{y=x p estremi}
\hat{t}_{21}^{(inc)}(\hat{x}_{(p)}) = \displaystyle n_l s_m \mathbb{K}_{lmkg} n_i s_j  \Delta_0 \int_{0}^1 \dot{t}_{ij,k}^g (\hat{x}_{(0)}+\zeta \Delta_0, \hat{x}_{(p)}) \, \salto{0.38}{\hat{v}}_{(1)}\sqrt{\zeta} \, d\zeta+\\
+\displaystyle n_l s_m \mathbb{K}_{lmkg} n_i s_j  \Delta_1 \int_{0}^1 \dot{t}_{ij,k}^g (\hat{x}_{(1)}+\zeta \Delta_1, \hat{x}_{(p)}) (\, \salto{0.38}{\hat{v}}_{(1)} (1-\zeta)+ \salto{0.38}{\hat{v}}_{(2)}\zeta \,) \, d\zeta+\\
+\displaystyle n_l s_m \mathbb{K}_{lmkg} n_i s_j \sum_{q=2}^{Q-3} \Delta_q \int_{0}^1 \dot{t}_{ij,k}^g (\hat{x}_{(q)}+\zeta \Delta_q, \hat{x}_{(p)}) (\, \salto{0.38}{\hat{v}}_{(q)} (1-\zeta)+ \salto{0.38}{\hat{v}}_{(q+1)}\zeta \,) \, d\zeta+\\
+\displaystyle n_l s_m \mathbb{K}_{lmkg} n_i s_j \Delta_{Q-2} \int_{0}^1 \dot{t}_{ij,k}^g (\hat{x}_{(Q-2)}+\zeta \Delta_{Q-2}, \hat{x}_{(p)}) (\, \salto{0.38}{\hat{v}}_{(Q-2)} (1-\zeta)+ \salto{0.38}{\hat{v}}_{(Q-1)}\zeta \,) \, d\zeta+\\
+\displaystyle n_l s_m \mathbb{K}_{lmkg} n_i s_j  \Delta_{Q-1} \int_{0}^1 \dot{t}_{ij,k}^g (\hat{x}_{(Q-1)}+\zeta \Delta_{Q-1}, \hat{x}_{(p)})\, \salto{0.38}{\hat{v}}_{(Q-1)} \sqrt{1-\zeta} \, d\zeta.
\end{multline}
When $p=1$, the finite parts of singular integrals can be evaluated as
\begin{multline*}
	\displaystyle  \Delta_0 \int_{0}^1 \dot{t}_{ij,k}^{g (st)} (\hat{x}_{(0)}+\zeta \Delta_0, \hat{x}_{(p)}) \, \salto{0.38}{\hat{v}}_{(1)}\sqrt{\zeta} \, d\zeta+\\	\displaystyle+ \Delta_1 \int_{0}^1 \dot{t}_{ij,k}^{g (st)} (\hat{x}_{(1)}+\zeta \Delta_1, \hat{x}_{(p)}) \, \salto{0.38}{\hat{v}}_{(1)} (1-\zeta) \, d\zeta=
\end{multline*}
\beq
	\displaystyle = T^g_{ijk}(\theta)\left(-\frac{9}{8 \Delta_{0}}-\frac{\ln{\Delta_{0}}}{2 \Delta_{0}}-\frac{1}{ \Delta_{1}}-\frac{\ln{ \Delta_{1}}}{ \Delta_{1}}\right)\,  \salto{0.38}{\hat{v}}_{(1)}, 
\eeq
while, when $p=Q-1$, the finite parts of the singular integrals can be evaluated as
\begin{multline*}
	\displaystyle  \Delta_{Q-2} \int_{0}^1 \dot{t}_{ij,k}^{g (st)} (\hat{x}_{(Q-2)}+\zeta \Delta_{Q-2}, \hat{x}_{(p)}) \, \salto{0.38}{\hat{v}}_{(Q-1)} \zeta\, d\zeta+\\\displaystyle+ \Delta_{Q-1} \int_{0}^1 \dot{t}_{ij,k}^{g (st)} (\hat{x}_{(Q-1)}+\zeta \Delta_{Q-1}, \hat{x}_{(p)}) \, \salto{0.38}{\hat{v}}_{(Q-1)} \sqrt{1-\zeta} \, d\zeta=
\end{multline*}
\beq
	\displaystyle = T^g_{ijk}(\theta)\left(-\frac{9}{8 \Delta_{Q-1}}-\frac{\ln{\Delta_{Q-1}}}{2 \Delta_{Q-1}}-\frac{1}{ \Delta_{Q-2}}-\frac{\ln{ \Delta_{Q-2}}}{ \Delta_{Q-2}}\right)\,  \salto{0.38}{\hat{v}}_{(Q-1)}.
\eeq
Hence, using a collocation method, thus assuming $p=1,\dots,Q-1$, a system of $Q-1$ algebraic equations is obtained which can be written in matrix form as follows
\beq
\label{linearsystem}
\left\{\hat{\bt}^{(inc)}_{21} \right\}=[\bC] \left\{\salto{0.38}{\hat{\bv}}\right\}.
\eeq

The nominal shear traction $\hat{t}^{(inc)}_{21}$ generated by a shear wave impinging the shear band can be obtained using equations 
(\ref{stream}) and (\ref{traction}) into equation (\ref{ondazza}), thus yielding
\beq 
\begin{split} \lb{t21hat} \hat{\dot{t}}^{(inc)}_{21}(\bx)=\tau_0 \,e^{i\frac{\Omega}{c}( \bp \cdot \bx - c t)}\bigl[&(n_1^2(\eta-1)-(1-k)n_2^2)\cos^2\theta_0+\\
 &+(n_2^2(\eta-1)+(1+k)n_1^2)\sin^2\theta_0+\\
 &+n_1n_2(\eta-2\xi)\sin 2\theta_0 \bigr].
\end{split} 
\eeq
where $\tau_0=i A \mu \Omega/c$ is the maximum shear stress acting at the shear wave front in the quasi-static limit, $\Omega\rightarrow0$. 
For a wave traveling orthogonally to the shear band, equation (\ref{t21hat}) reduces to a positive quantity, at least until strong ellipticity holds true.

\section{Results for the $J_2$-deformation theory of plasticity}

A shear band is discretized with $Q=100$ line elements and numerically analyzed when inside a ductile metal whose behaviour is described by the $J_2$-deformation theory of plasticity. The incremental moduli are 
provided by equations (\ref{modulazzicertamente}) and the hardening exponent is assumed to be $N=0.4$, so that ellipticity is lost at 
the critical value of the logarithmic strain $\varepsilon_1 \approx 0.678$. Results are presented below.

\subsection*{Wave propagation normal to the shear band}

The direction of the wave propagation is now considered to be orthogonal to the shear band faces, so that the whole front of the band is uniformly loaded.
The numerical solution of the linear system (\ref{linearsystem}) allows to compute the longitudinal displacement jump across the shear band, $\salto{0.1}{\hat{v}_1}$. 

\begin{figure}[!h]
	\begin{center}
		\includegraphics[width=1\textwidth]{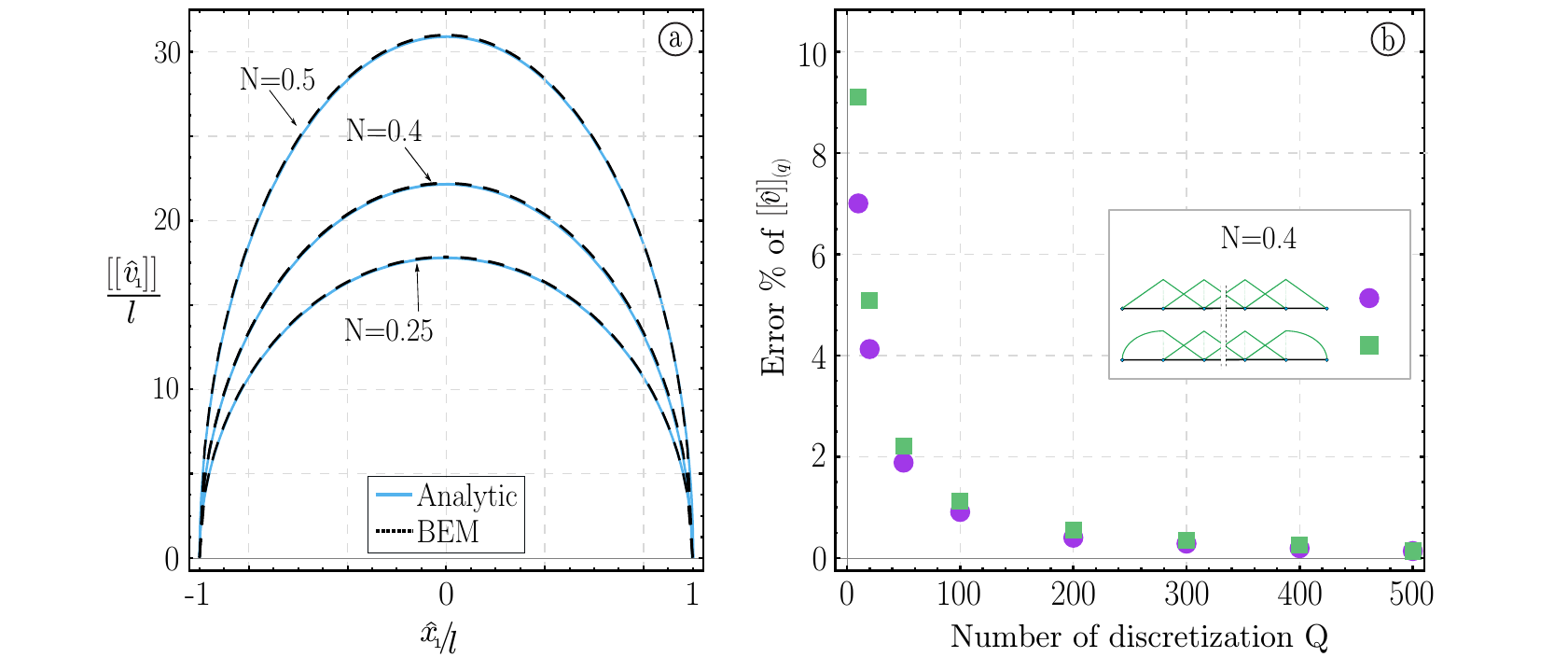}
		\caption{\footnotesize The quasi-static behaviour of a shear band loaded with a remote shear (obtained numerically in the limit $\Omega \rightarrow 0$) is compared with an available analytical solution 
		for different hardening exponents $N$ and prestrains near the elliptic border. Part (a): Modulus of dimensionless displacement jump along the shear band line, $\hat{x}_1/l$, for the $J_2$-deformation theory of plasticity and three hardening exponents $N \,(0.25,0.4,0.5)$. 
			Part (b):	Percent error in the incremental displacement jump $\salto{0.38}{\hat{v}}_{q}$ for different numbers of collocation points $Q \, (10,20,50,100,200,500)$, and for two sets of shape functions ($N=0.4$ has been considered). The errors are evaluated at the middle of the shear band, $\hat{x}_1/l=0$, note that the circular and square spots are practically superimposed for $Q$>200.}
		\label{err}
	\end{center}
\end{figure}

A validation of the developed numerical technique can be obtained, in the limit $\Omega\rightarrow 0$, by comparing with the analytic solution for the static case provided by Bigoni and Dal Corso \cite{bigonifdc}.
This validation is provided in Figure \ref{err}(a), where 
the modulus of the displacement jump $\salto{0.1}{\hat{v}_{1}}$ (divided by the semilength of the shear band) is plotted along the shear band line $\hat{x}_1$. 
The validation turns out to be excellent, as the analytic solution is superimposed to the numerical solutions, 
for different values of the hardening exponent $N$ ($0.25, 0.4, 0.5$), at respective levels of prestrain close to the elliptic boundary ($\varepsilon_1=0.522, \varepsilon_1=0.667, \varepsilon_1=0.771$).

The convergence of the numerical solution to the static -analytical- solution (developed in \cite{bigonifdc}) is shown in Figure 
\ref{err}(b), where the (percent) error in the incremental displacement jump $\salto{0.38}{\hat{v}}_{q}$, evaluated at the middle of the shear band, $\hat{x_1}/l=0$, is reported as a function of the number of the collocation points $Q$. 
Two different sets of shape functions are considered, namely, linear shape functions for the whole shear band in one case (circular spots), while in the other case square-root shape functions are used only in the element at the shear band tip (square spots). It can be seen that in the middle of the shear band for $Q=100$ the error is about $1\%$ for both shape function sets.    

With $Q=100$ elements and the selected shape functions, the computing time necessary to find the displacement jump across the shear band ranges between 15 and 20 minutes 
running the software Mathematica 11.2
in a computer AMD Opteron cluster Stimulus (available at the \lq Instability Lab' of the University of Trento).

\begin{figure}[!h]
  \begin{center}
    \includegraphics[width=1\textwidth]{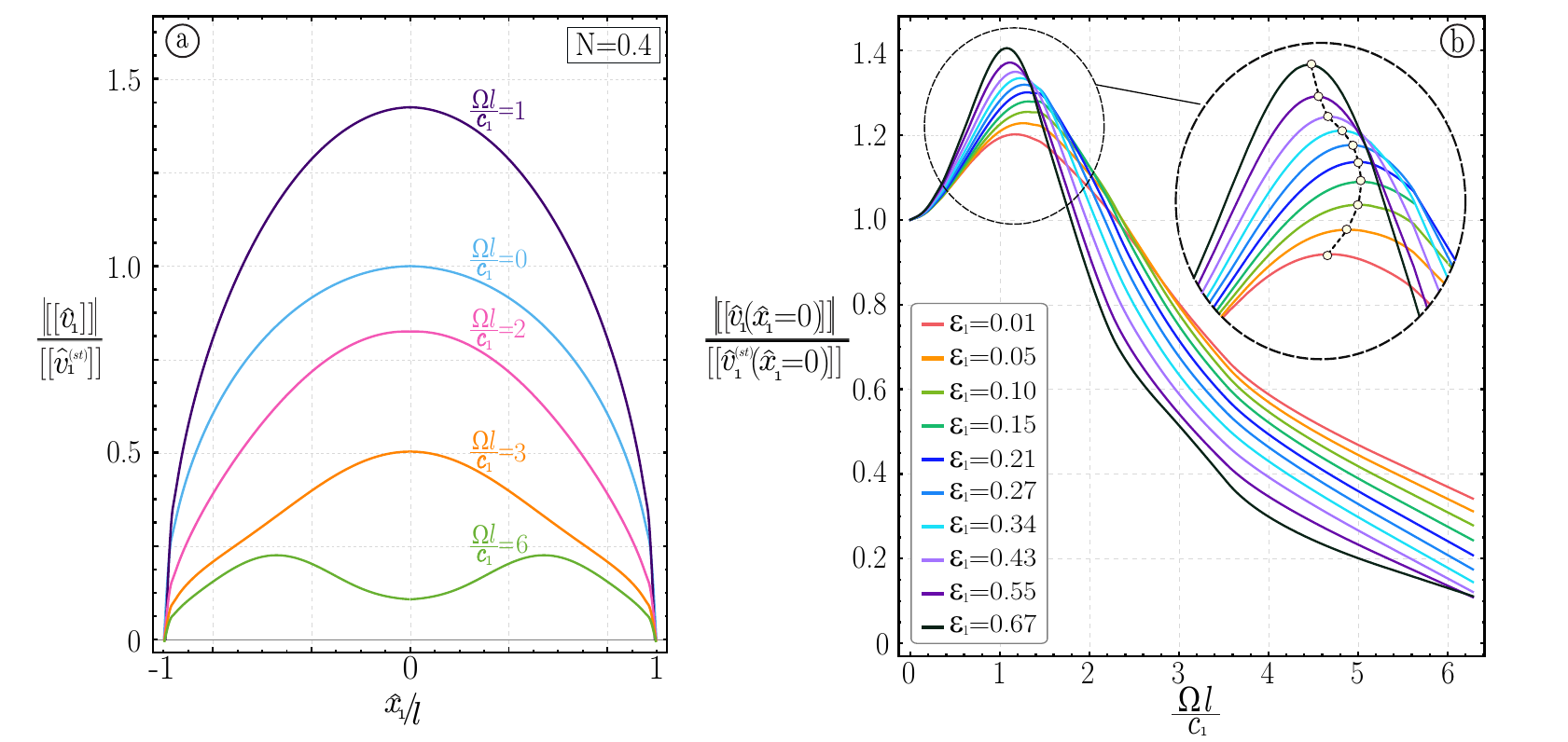}
    \caption{\footnotesize 
    Modulus of dimensionless displacement jump along the shear band line, $\hat{x}_1/l$, for the $J_2$-deformation theory of plasticity. Part (a): different wavenumber are considered with $N=0.4$ and prestrain $\varepsilon_1=0.667$.
    Part (b): the dimensionless displacement jump in the middle of the shear band ($\hat{x}_1=0$) is plotted as a function of the dimensionless frequency for different values of prestrain with $N=0.4$. 
    Note that a resonance frequency is visible (the peak of the curves) and that this resonance becomes more evident at increasing prestrain, when it approaches the elliptic boundary.}
    \label{n04}
  \end{center}
\end{figure}

The dynamic shape of the displacement jump along the shear band line is reported in Figure \ref{n04}(a), referred to a prestrain $\varepsilon_1=0.667$, close to the boundary of ellipticity loss.   
This figure shows that, near the resonance frequency, the displacement jump along the shear band assumes the quasi-static shape, but at high frequency displays a markedly different behaviour \cite{mal}, namely, it  decades in amplitude and displays an oscillation (see the green curve referred to $\Omega l/c_1=6$).

The variation with the wavenumber, $\Omega l/c_1$, (of the modulus) of the displacement jump $\salto{0.1}{\hat{v}_{1}}$ (normalized with respect to the quasi-static value $\salto{0.1}{\hat{v}_{1}^{(st)}}$) is shown in Figure \ref{n04}(b) for several values of prestrain, ranging from 0 to $\varepsilon_1=0.667$. 
In this figure the maxima of the curves represent resonance condition (the displacement grows, but does not blow-up to infinity, due to the radiation damping, properly accounted for in the numerical solution), so that it is clear that an increase in the prestrain leads to an amplification factor which grows from $20\%$, occurring at null prestrain, to $41\%$, occurring at a prestrain close to the border of ellipticity loss. Results not reported for brevity show that a decrease in the hardening exponent $N$ shifts the resonance towards higher frequencies.

\begin{figure}[!h]
	\begin{center}
		\includegraphics[width=1\textwidth]{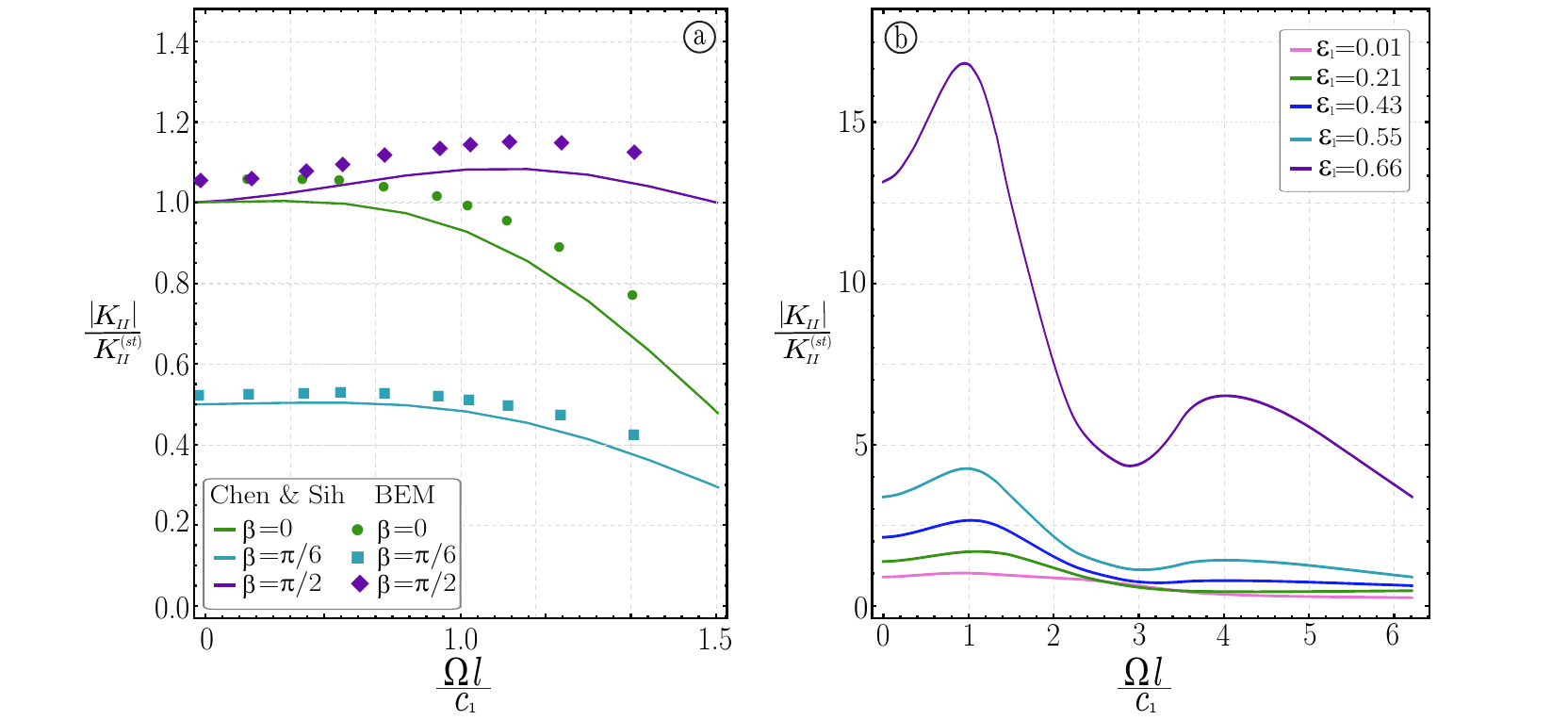}
		\caption{\footnotesize Modulus of dimensionless mode II Stress Intensity Factor at the shear band tip as a function of the wavenumber: (a) comparison with the analytical solution of Chen and Sih \cite{chen}, with null prestrain in the isotropic case with $\mu=\mu_{*}$, (b) for different levels of prestrain for a $J_2$ deformation theory with $N=0.4$.}
		\label{sif}
	\end{center}
\end{figure}

The stress concentration at the shear band tips can be investigated using the Stress Intensity Factor (SIF) and because only incremental shear stresses are acting on the band, a Mode II SIF is adopted, which is defined as
\beq K_{II}=\lim_{\hat{x}_1 \to l^+} \hat{\dot{t}}_{21}(\hat{x}_1,\hat{x}_2=0)\sqrt{2 \pi (\hat{x}_1-l)}
\eeq
which in the quasi-static case becomes 
\beq K_{II}^{st}=\hat{\dot{t}}_{21}^\infty \sqrt{ \pi\,l}.
\eeq
The SIF is also defined as a function of the displacement jump in the form \cite{barra,chirino}
\beq K_{II}=\frac{\mu \sqrt{2 \pi}}{4(1-\nu)}\frac{\salto{0.1}{\hat{v}}_{(1)}}{\sqrt{\Delta_0}} ,
\eeq
where $\salto{0.1}{\hat{v}}_{(1)}$ is the displacement jump evaluated at the first inner node from the tip of the shear band. 
Figure \ref{sif} reports the SIF, $K_{II}$, normalized through the quasi-static condition $K_{II}^{(st)}$, as a function of the of the wavenumber.

%%%%%%%%%%%%%%%%%%%%%%%%%%%%%%%%%%%%%%%%%%%%%%%%%%
%%%%%%%%%%%%%%%%%%%%%%%%%%%%%%%%%%%%%%%%%%%%%%%
It has to be noted that Chen and Sih \cite{chen} developed an analytical solution for the SIF pertinent to a crack impinged by an incident shear wave in a linear elastic and isotropic body. This solution can be used to validate the developed numerical procedure, as reported in part (a) of Figure \ref{sif}, relative to a null prestrain. Here the absolute 
value of the SIF (normalized with respect to the quasi-static limit) is reported as a function of the wavenumber. The validation turns out to be satisfactory, because for the tested angles $\beta$ of the wave propagation, the discrepancy is within $8\%$.

The dimensionless SIF for the shear band tips at different levels of prestrain is reported in Figure \ref{sif}(b) as a function of the dimensionless frequency. In the quasi-static limit and for a null prestrain the SIF 
correctly tends to 1, while, when the elliptic boundary is approached, the SIF blows up, reaching a value approximately 15 times the quasi-static value for a prestrain $\varepsilon_1=0.66$, whereas at the elliptic border it grows to infinity, coherently with the quasi-static behaviour, \cite{bigonifdc}. This is once more the evidence of a resonance condition, with an increase of $41\%$ of the SIF with respect to the quasi-static case.

It is important to remark that both evidences presented in Fig. \ref{n04}(c) and \ref{sif}(b) show that the presence of a shear band 
produces a resonance, evidenced through a substantial growth in the jump of displacement across the shear band and in the stress intensity factor at the 
shear band tip.

\subsection*{Wave propagation inclined or parallel to the shear band}

A wave obliquely impinging a shear band is now considered, with $\bp \scalp \bn$ different from both 0 and 1. The shear traction can be 
derived from equation (\ref{t21hat}) and is composed of a real symmetric part and an imaginary skew-symmetric part. Therefore, the traction is non-symmetric with respect to the $\hat{x}_2$-axis. 
\begin{figure}[!h]
  \begin{center}
    \includegraphics[width=0.95\textwidth]{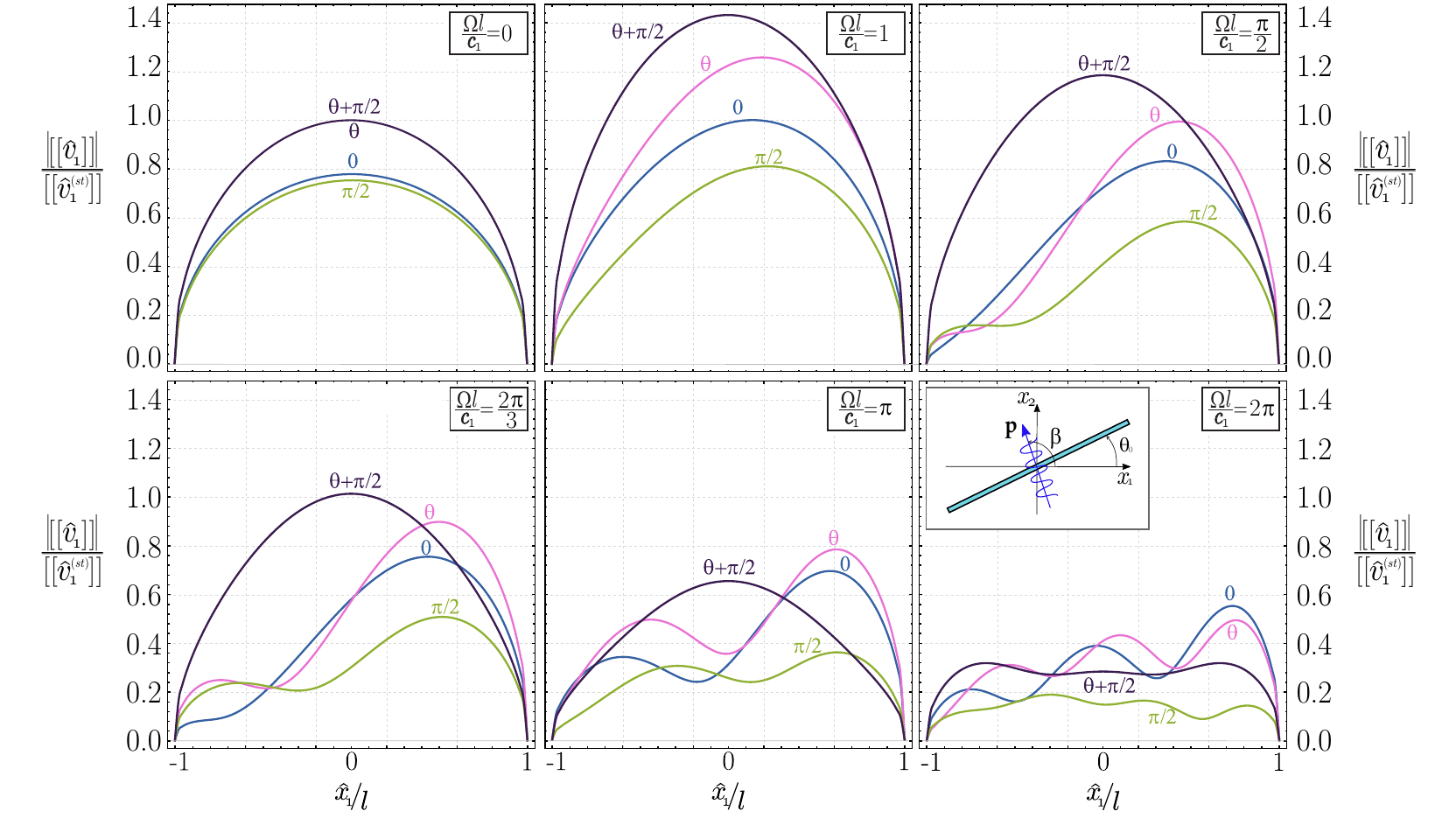}
    \caption{\footnotesize Modulus of dimensionless displacement jumps (for a $J_2$-deformation theory of plasticity material with $N=0.4$) along the shear band line, $\hat{x}_1/l$, at various wavenumber. For each wavenumber, four different inclinations $\beta$ of the wave propagation are considered ($0, \theta_0, \pi/2, \pi/2+\theta_0$). }
    \label{skw}
  \end{center}
\end{figure}

This can be noted in Figure \ref{skw}, where, as in Fig. \ref{n04}(b), the dimensionless displacement jump is reported along the shear band line as a function of the dimensionless frequency, for various inclinations of the wave propagation vector. When the wave propagation is inclined at an angle $\beta$ belonging to the interval $(- \pi/2+\theta_0, \pi/2+\theta_0)$, the maximum value of the displacement jump shifts towards the right tip of the band. 

\begin{figure}[!h]
  \begin{center}
    \includegraphics[width=1\textwidth]{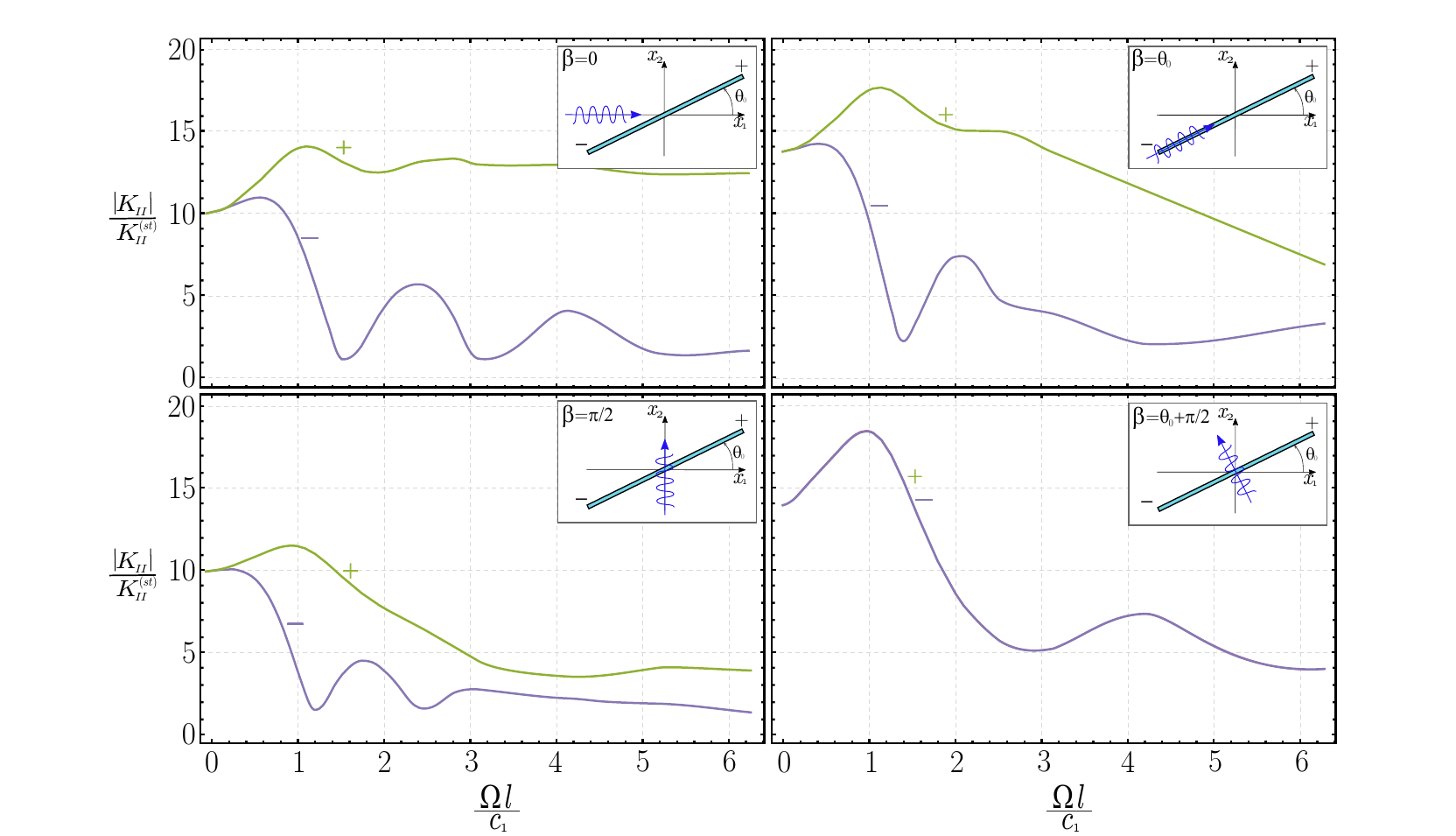}
    \caption{\footnotesize SIF at the left \lq $-$' and right \lq $+$' tips of a shear band (in a $J_2$-deformation theory of plasticity material with $N=0.4$), for different inclinations $\beta$ of the wave propagation ($0, \theta_0, \pi/2, \pi/2+\theta_0$). }
    \label{skwsif}
  \end{center}
\end{figure}

Due to the fact that the wave is now inclined with respect to the shear band, the stress intensity factors at the tips of the shear band are different \cite{vanderhijiden}, see Figure \ref{skwsif}, where the dimensionless SIF for the two tips (one denoted by \lq +' and the other by \lq $-$') are reported as functions of the dimensionless frequency. It can be observed that the higher the displacement jump, the higher is the SIF, moreover a wave orthogonal to the shear band produces the largest value of the SIF and therefore the maximum resonance. 

\subsection*{Incremental strain fields}

The modulus of the incremental strain field (which is deviatoric, because of incompressibility) defined as
$
\left( v_{i,j} v_{i,j} + v_{i,j}v_{j,i}\right)/2,
$
can be computed by using the gradient of the incremental displacement, equation (\ref{scamorza3}), in the constitutive equations (\ref{traction}). In the following the modulus of the
incremental strain field is computed by using the real part of the gradient of incremental displacement, so that the phase shift related to the imaginary part is not considered. 

\begin{figure}[!h]
  \begin{center}
    \includegraphics[width=1\textwidth]{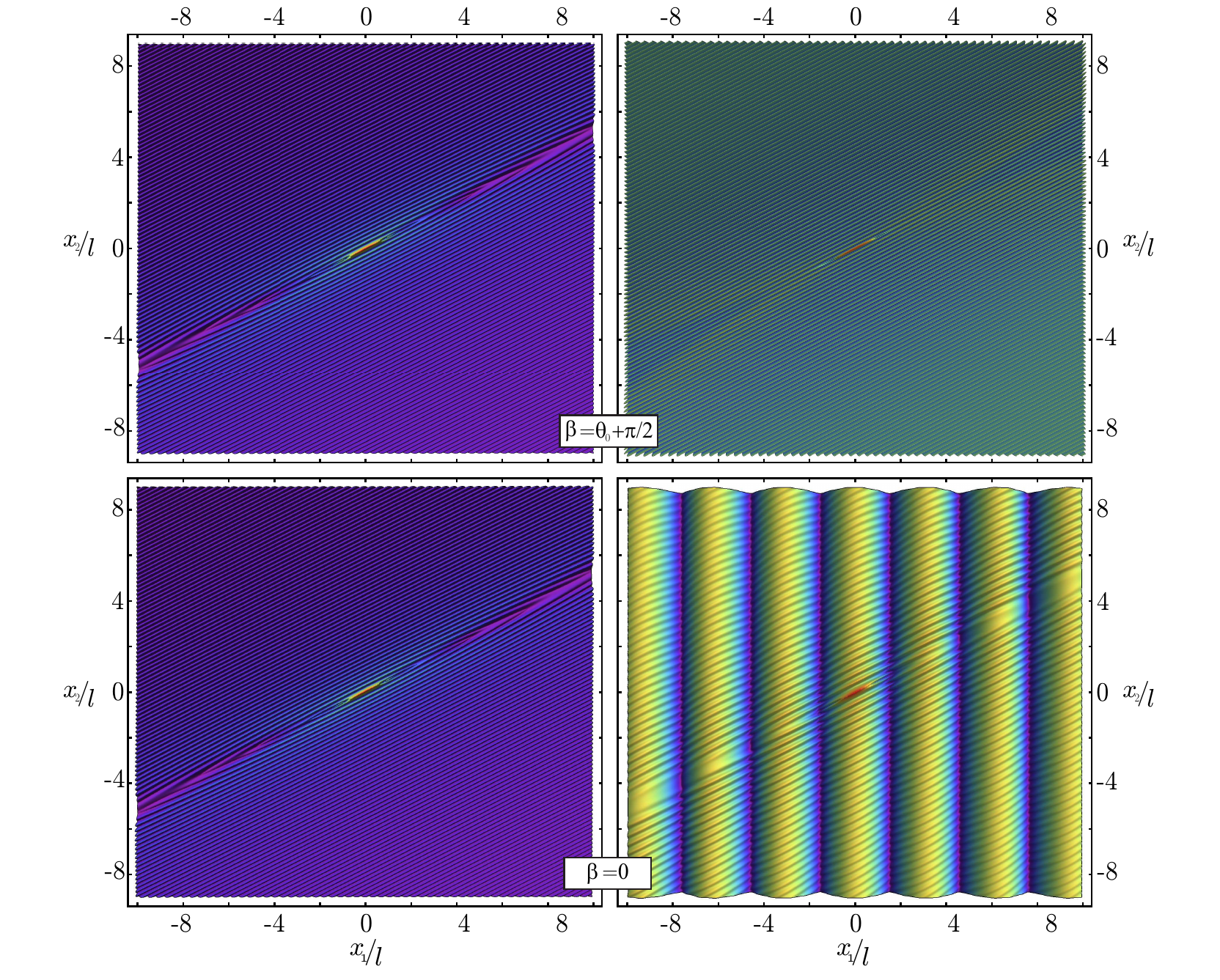}
    \caption{\footnotesize Scattered (left) and total (right) incremental strain field produced by a wave incident to a shear band (in a $J_2$-deformation theory of plasticity material with $N=0.4$) orthogonally to it ($\beta=\theta_0+\pi/2$) or aligned parallel to the $x_1-$axis ($\beta=0$). The wavenumber is $\Omega l/c_1=1$ and the level of prestrain is $\varepsilon_1=0.667$, close to the elliptic boundary.}
    \label{field1}
  \end{center}
\end{figure}

The modulus of the incremental strain field, computed at a prestrain level of $\varepsilon_1=0.667$ (i.e. close to the elliptic boundary) is reported in Figure \ref{field1}, in terms of scattered wave field (on the left) and in terms of total wave field (on the right). Two incident waves with wavenumber $\Omega l/c_1=1$ are considered, one orthogonal to the shear band (with inclination $\beta=\theta_0+\pi/2$) and the other aligned parallel to the $x_1-$axis (with inclination $\beta=0$). These inclinations of propagation represent the directions along which the wave velocity $c$ assumes the minimum and maximum values respectively, see equation (\ref{calfa}).

It is worth noting that the wavenumber $\Omega l/c_1=1$ used for the computations corresponds to a wavelength in the direction orthogonal to the shear band, $2\pi l c/c_1$, which is approximately 1/6 of the shear band length, thus much greater than the shear band thickness. 

\begin{figure}[!h]
	\begin{center}
		\includegraphics[width=1\textwidth]{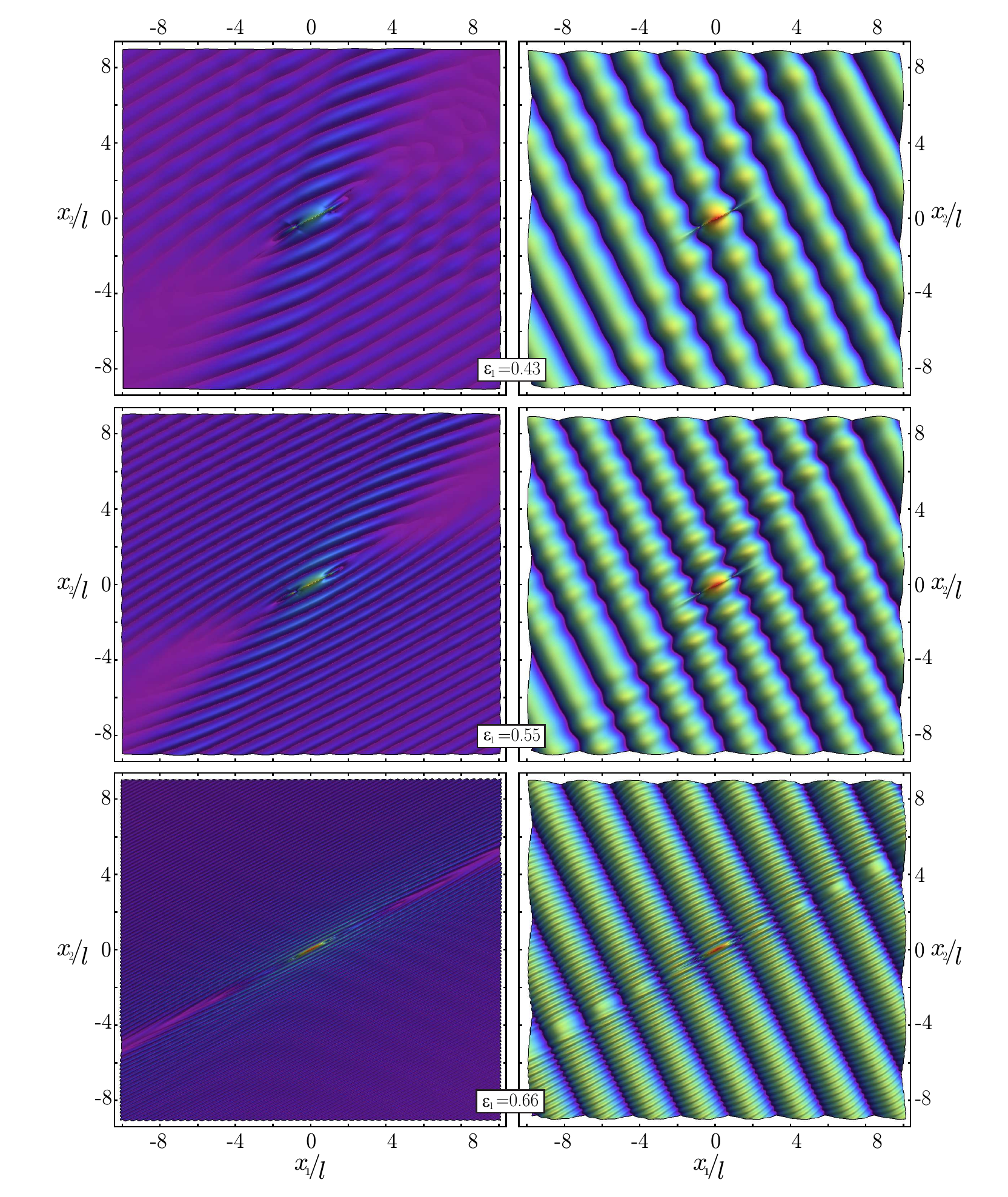}
		\caption{\footnotesize Scattered (left) and total (right) incremental strain field produced by a wave impinging parallel to a shear band, $\beta=\theta_0$, (in a $J_2$-deformation theory of plasticity material with $N=0.4$). Various levels of prestrain $\varepsilon_1=0.43, \varepsilon_1=0.55,\varepsilon_1=0.66$ are reported with wavenumber $\Omega l/c_1=1$.}
		\label{field2}
	\end{center}
\end{figure}

It can be noted that for both wave inclinations, the scattered field turns out to be a family of plane waves parallel to the shear band. 
The effect of this scattered field on the total strain field is to produce a fine texture of parallel vibrations, which superimposes on the impinging wave field. The texture shows a narrow conical shadow zone emanating from the shear band tips, where the scattered field is strongly attenuated and tends to disappear. This effect becomes more visible in the case of $\beta=0$, because incident and scattered waves propagate in different directions, rather than in the case of wave travels orthogonal to the band.

In the case of an incident wave with wavenumber $\Omega l/c_1=1$, propagating in the direction parallel to the shear band, Figure \ref{field2} represents the scattered and total  strain fields for three increasing levels of prestrain  ($\varepsilon_1=0.43, \varepsilon_1=0.55, \varepsilon_1=0.66$). Starting from the lowest level of prestrain, the unperturbed conical zone is already visible, but this zone tends to become narrower when the elliptic boundary is approached.  

\begin{figure}[!h]
	\begin{center}
		\includegraphics[width=1\textwidth]{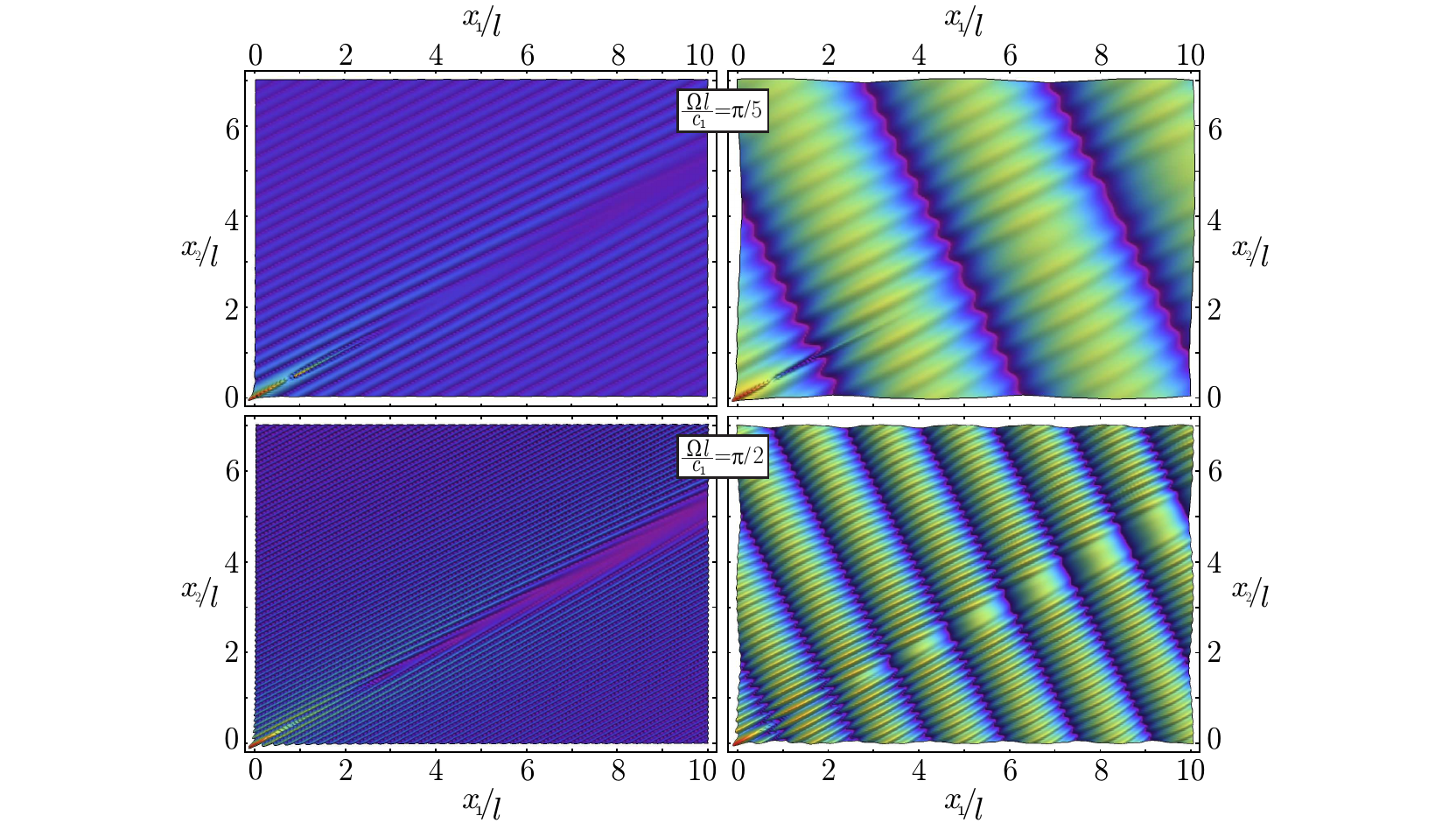}
		\caption{\footnotesize Incremental strain field near a shear band (in a $J_2$-deformation theory of plasticity material with $N=0.4$) produced by a wave impinging parallel to the shear band, $\beta=\theta_0$ and waveleght $\Omega l/c_1=\pi/5$ (upper part) and $\Omega l/c_1=\pi/2$ (lower part).}
		\label{fieldpenn}
	\end{center}
\end{figure}

The shadow zone is analyzed near the elliptic boundary as a function of the frequency, in particular, the upper left quarter of the map of the incremental strain field is reported in Figure \ref{fieldpenn} for two frequencies 
($\Omega l/c_1= \pi/5,\Omega l/c_1 = \pi/2 $). This plot reveals that the shadow zone becomes more visible at frequencies higher than the value corresponding to resonance.
 
It is remarked that experimental results on time-harmonic vibration of shear bands are not available in the literature. 
However, a close scrutiny of experiments on dynamic propagation of shear bands induced by an impact loading 
reveals that waves propagate in the material with the shear band inclination (\cite{lis}, their figures 4, 5, 6, and 8), a feature 
also observed in our simulations (Figures  \ref{field2} and \ref{fieldpenn}). Moreover, results relative to wave propagation in highly orthotropic materials 
exhibit shadow zones similar to those visible in Figure \ref{fieldpenn} (\cite{Shen}, their figure 3).
Finally, experiments on wave propagation in an aluminum solid containing a crack with attritive faces, which are prestressed in compression, (\cite{Blanloeuil}, their figure 8)
can be compared with results pertinent to Mooney-Rivlin material presented in Appendix B.  In both cases waves parallel to the discontinuity line are evidenced.

\section{Conclusions}

The dynamic effects induced by a shear band formed in a material strained close to the failure of ellipticity can be analyzed 
through the development of {\it ad hoc} boundary integral equations and collocation techniques. Results show that the shear band 
produces a complex dynamic interaction with impinging waves generating resonance at a certain frequency. This resonance promotes shear band development, as can be revealed by the 
variation with frequency of the stress intensity factor at the shear band tips. Moreover, the vibration pattern generated near the shear band shows a fine development of plane waves and the formation of a narrow zone 
of low incremental strain emerging from the shear band edges and propagating with a conical shape. 
It is worth noting that the results obtained in the present article can be generalized in several ways. In particular: (i.) three-dimensional problems can be analyzed in which the shear band can assume a complex form, for instance penny-shaped, conical or curved; (ii.) materials with different constitutive equations and even materials different from 
metals, for instance granular matter, can be considered; (iii.) transient dynamics may be studied. In all these cases, the boundary integral equations developed in this article are either still valid or require minimal modifications, so that only the Green's function has to be determined,
a difficulty which can be attacked with the methods shown in \cite{cono, picc, willis}.

\section*{Acknowledgement}
Financial support from the ERC advanced grant ERC-2013-ADG-340561-INSTABILITIES is gratefully acknowledged.

% \clearpage
%%%%%%%%%%%%%%%%%%%%%%%%%%%%%%%%%%%%%%%%%%%%%%%%%%%%%%%%%%%%%%%%%%%%%%%
%Appendix
\appendix
\renewcommand{\theequation}{\thesection.\arabic{equation}}

% \begin{center}
% \bf \Large Appendix
% \end{center}

\section{Appendix A}

\setcounter{equation}{0}

The gradient of the nominal stress tensor $\dot{t}^g_{ij,k}$ is obtained by the plane wave expansion as 
\beq 
\label{planewaveexp}
\dot{t}^g_{ij,k}=-\frac{1}{4\pi^2} \int_{\bomega} \tilde{t}^g_{ij,k}(\bomega)\, d \omega ,
\eeq
and it can be decomposed into a static and a dynamic contribution as
\beq
\tilde{t}^g_{ij,k}(\bomega)=\tilde{t}^{g (st)}_{ij,k}(\bomega)+\tilde{t}^{g (dyn)}_{ij,k}(\bomega),
\eeq
which are defined as
\beq
\lb{sttt}
\tilde{t}^{g (st)}_{ij,k}(\bomega)=-\frac{\tilde{F}^g_{ijk}(\bomega)}{(\bomega \scalp \bx)^2},
\eeq
\beq
\tilde{t}^{g (dyn)}_{ij,k}(\bomega)=\left[\omega_g \omega_k \delta_{ij}- \tilde{F}^g_{ijk}(\bomega)\right] \eta^2 \Xi'(\eta \bomega \scalp \bx),
\eeq
where a prime denotes differentiation with respect to $\bomega\scalp\bx$ and 
\begin{multline}
\tilde{F}^g_{ijk}(\bomega)=\forK_{ijhl} \omega_k \omega_h \frac{\delta_{lg}-\omega_l \omega_g}{L(\bomega)}+\left(\frac{\omega_g(2 \mu_*-\mu)(1-\omega_g^2)}{L(\bomega)}\right.+\\\left.+\frac{(\mu-(\delta_{2g}-\delta_{1g}))\frac{\sigma}{2} \omega_g^2}{L(\bomega)}
-\frac{\sigma}{2}\omega_k\frac{\omega_1\omega_2^2}{L(\bomega)}\right) \delta_{ij}.
\end{multline}
By specifying the unit vector $\bomega$ as
\beq
\omega_1=\cos (\alpha+\theta), \quad \omega_2=\sin (\alpha+\theta),
\eeq
the static contribution (\ref{sttt}) is strongly singular in both the variables $\alpha$ and $r$, so that, to evaluate the plane wave expansion (\ref{planewaveexp}), 
the quantity $\tilde{t}^{g (st)}_{ij,k}$ is regularized as follows 
\beq
\tilde{t}^{g (st)}_{ij,k}(\bomega)=\frac{\tilde{T}^g_{ijk}\left(\alpha,\theta\right)}{r^2} ,
\eeq
where 
\beq
\tilde{T}^g_{ijk} = 
-\frac{1}{\cos^2{\alpha}} \left[\tilde{F}^g_{ijk}(\alpha+\theta)-\left(\tilde{F}^g_{ijk}(\alpha+\theta)\right)_{\alpha=\pi/2}-\left(\alpha-\frac{\pi}{2}\right) \left(\tilde{F}^g_{ijk}(\alpha+\theta)\right)^{'}_{\alpha=\pi/2}\right], 
\eeq
in which the prime denotes differentiation with respect to $\alpha$. 

\section{Appendix B}

\setcounter{equation}{0}

\begin{figure}[!h]
	\begin{center}
		\includegraphics[width=1\textwidth]{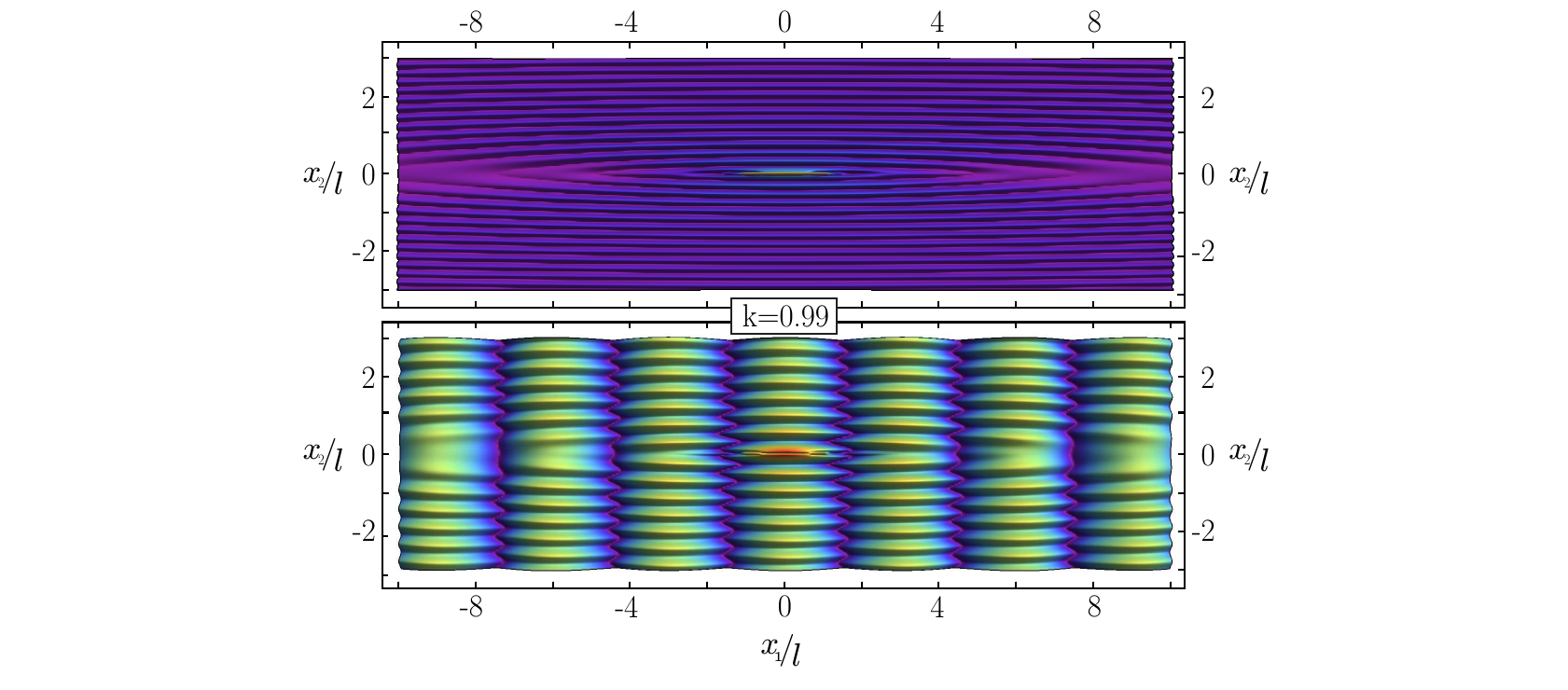}
		\caption{\footnotesize Scattered (upper part) and total (lower part) incremental strain field produced by a wave impinging parallel to a shear band, $\beta=\theta_0$, in a Mooney-Rivlin material. A wavenumber $\Omega l/c_1=1$ and a prestress $k=0.99$ (near the elliptic/parabolic boundary) have been considered.}
		\label{mooney}
	\end{center}
\end{figure}

With reference to a Mooney-Rivlin material, the constitutive equation in plane strain condition is always given in the form (\ref{traction}), where $\mu$ and $\mu_{*}$ coincide and are expressed in terms of the in-plane stretches $\lambda \geq 1$ and $1/\lambda$ as 
\beq
\mu=\mu_{*}=\frac{\mu_0}{2}(\lambda^2+\frac{1}{\lambda^2}).
\eeq
A shear band in a Mooney-Rivlin material, emerges and grows parallel to the $\sigma_1$-principal axis, with a null inclination, $\theta_0=0$. In order to approach the 
elliptic/parabolic boundary, the limit of the prestress $k\rightarrow1$ (which formally corresponds to infinite stretch) is considered. \\
Due to the inclination $\theta_0=0$, the integral equation (\ref{pallissime}) simplifies to
\beq
\lb{pallissimeteta0}
\hat{t}_{21}^{(inc)}(\by) =  \int_{-l}^l \left[(\mu-p)\dot{t}^2_{21,1}(\hat{x}_1, \by)+\left(\mu-\frac{\sigma}{2}\right)\dot{t}^1_{21,2}(\hat{x}_1, \by) \right]\, \salto{0.38}{\hat{v}_1} \, d\hat{x}_1.
\eeq
Figure \ref{mooney} shows the scattered (upper part) and total (lower part) incremental deviatoric strain fields, for a shear band in a Mooney-Rivlin material at prestress $k=0.99$, near the elliptic boundary. An incident wave is considered travelling parallel to the shear band ($\beta=0$), with wavenumber $\Omega l/c_1=1$.
The emergence of plane waves, parallel to the shear band, can be noted and the formation of a conical shadow zone is visible, which can be compared with the experimental results on wave propagation in a rectangular aluminum block containing a crack prestressed through compressive forces orthogonal to the crack faces 
\cite{Blanloeuil} (their figure 8).

%\clearpage
 { \singlespace
}

%%%%%%%%%%%%%%%%%%%%%%%%%%%%%%%%%%%%%%%%%%%%%%%%%%%%%%%%%%%%%%%%%%%%%%%%%%%%%%%%%%%%%%%%%%%%%%%%%%%%%%%%

\end{document}